# Electrostatic Conversion for Vibration Energy Harvesting

S. Boisseau, G. Despesse, B. Ahmed Seddik

*LETI, CEA, Minatec Campus, Grenoble, France*

## Abstract

This chapter focuses on vibration energy harvesting using electrostatic converters. It synthesizes the various works carried out on electrostatic devices, from concepts, models and up to prototypes, and covers both standard (electret-free) and electret-based electrostatic vibration energy harvesters (VEH).

After introducing the general concept of Vibration Energy Harvesting and the global advantages and drawbacks of electrostatic devices to convert mechanical power into electricity (§1), we present in details the conversion principles of electret-free and electret-based electrostatic converters and equations that rule them in §2. An overview of electrostatic VEH, comparing the results from several laboratories (powers, sizes, concepts…) is provided in §3. In §4, we introduce several power management circuits dedicated to electrostatic VEH. These circuits are extremely important as they are the only way to turn VEH output powers into viable supply sources for electronic devices (sensors, microcontrollers, RF chips…). Assessments, limits and perspectives of electrostatic VEH are then presented in §5.

## 1. From thousands to millions sensors in our environment

"Everything will become a sensor"; this is a global trend to increase the amount of information collected from equipment, buildings, environments… enabling us to interact with our surroundings, to forecast failures or to better understand some phenomena. Many sectors are involved: automotive, aerospace, industry, housing. Few examples of sensors and fields are overviewed in Figure 1.

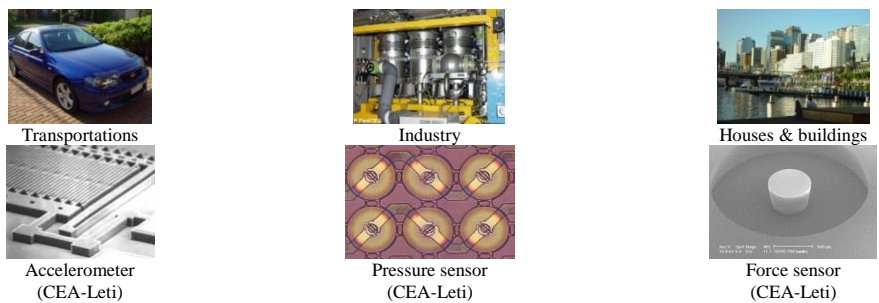

Transportations · Industry · Houses & buildings

Accelerometer (CEA-Leti) · Pressure sensor (CEA-Leti) · Force sensor (CEA-Leti)

Figure 1. Millions sensors in our surroundings

Unfortunately, it is difficult to deploy many more sensors with today's solutions, for two main reasons:

1. Cables are becoming difficult and costly to be drawn (inside walls, on rotating parts)
2. Battery replacements in wireless sensor networks (WSN) are a burden that may cost a lot in large factories (hundreds or thousands sensor nodes).

As a consequence, industrialists, engineers and researchers are looking for developing autonomous WSN able to work for years <u>without any human intervention</u>. One way to proceed consists in using a green and theoretically unlimited source: ambient energy [1].

## 1.1. Ambiant Energy & Applications

Four main ambient energy sources are present in our environment: mechanical energy (vibrations, deformations), thermal energy (temperature gradients or variations), radiant energy (sun, infrared, RF) and chemical energy (chemistry, biochemistry).

These sources are characterized by different power densities (Figure 2). Energy Harvesting (EH) from outside sun is clearly the most powerful (even if values given in Figure 2 have to be weighted by conversion efficiencies of photovoltaic cells that rarely exceed 20%). Unfortunately, solar energy harvesting is not possible in dark areas (near or inside machines, in warehouses). And similarly, it is not possible to harvest energy from thermal gradients where there is no thermal gradient or to harvest vibrations where there is no vibration.

As a consequence, the source of ambient energy must be chosen according to the local environment of the WSN's node: <u>no universal ambient energy source exists</u>.

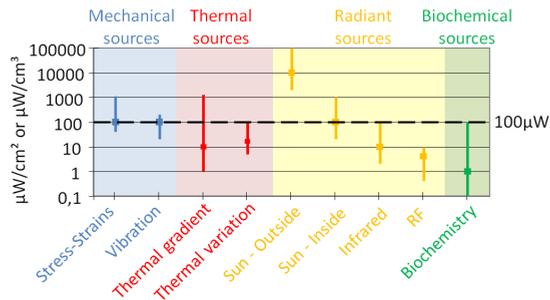

Figure 2. Ambient sources power densities before conversion

Figure 2 also shows that 10-100µW of available power is a good order of magnitude for a 1cm² or a 1cm³ energy harvester. Obviously, 10-100µW is not a great amount of power; yet it can be enough for many applications and especially WSN.

## 1.2. Autonomous Wireless Sensor Networks & Needs

A simple vision of autonomous WSN' nodes is presented in Figure 3(a). Actually, autonomous WSN' nodes can be represented as 4 boxes devices: (i) "sensors" box, (ii) "microcontroller (µC)" box, (iii) "radio" box and (iv) "power" box. To power this device by EH, it is necessary to adopt a "global system vision" aimed at reducing power consumption of sensors, µC and radio.

Actually, significant progress has already been accomplished by microcontrollers & RF chips manufacturers (Atmel, Microchip, Texas Instruments…) both for working and standby modes. An

Electrostatic Conversion for Vibration Energy Harvesting

example of a typical sensor node's power consumption is given in Figure 3(b). 4 typical values can be highlighted:

- 1-5µW: µC standby mode's power consumption
- 500µW-1mW: µC active mode's power consumption
- 50mW: transmission power peak
- 50-500µJ: the total amount of energy needed to perform a complete measurement and its wireless transmission, depending on the sensor and the RF protocol.

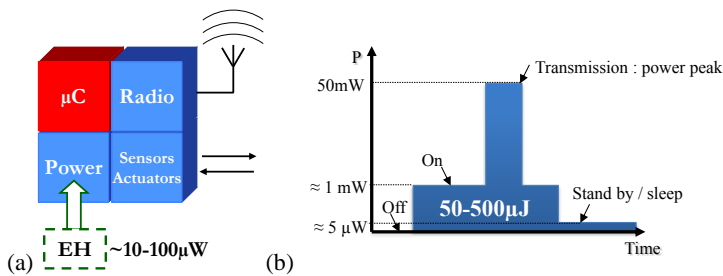

Figure 3. (a) Autonomous WSN node and (b) sensor node's power consumption

Then, the energy harvester has to scavenge at least 5µW to compensate the standby mode's power consumption, and a bit more to accumulate energy (50-500µJ) in a storage that is used to supply the following measurement cycle.

Today's small scale EH devices (except PV cells in some cases) cannot supply autonomous WSN in a continuous active mode (500µW-1mW power consumption vs 10-100µW for EH output power). Fortunately, thanks to an ultra-low power consumption in standby mode, EH-powered autonomous WSN can be developed by adopting an intermittent operation mode as presented in Figure 4. Energy is stored in a buffer (a) (capacitor, battery) and used to perform a measurement cycle as soon as enough energy is stored in the buffer (b & c). System then goes back to standby mode (d) waiting for a new measurement cycle.

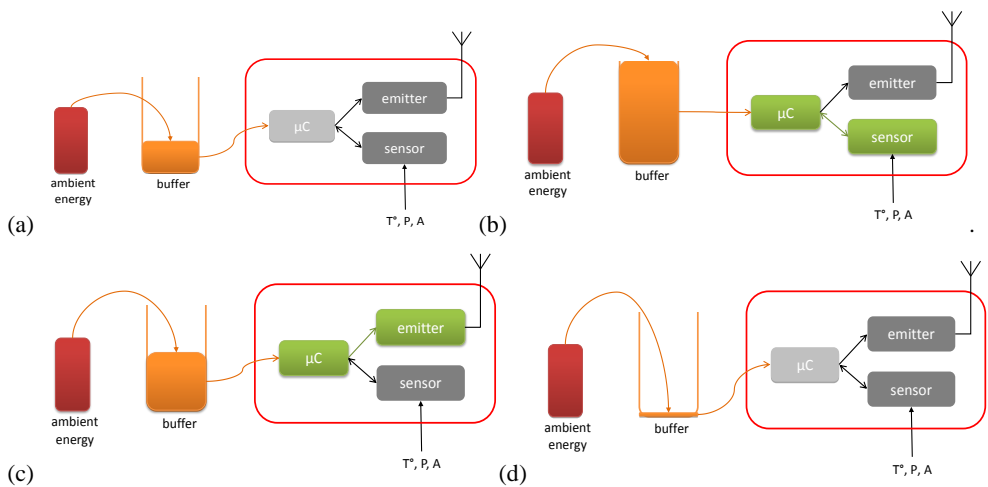

Figure 4. WSN measurement cycle

Therefore, it is possible to power any application thanks to EH, even the most consumptive one; the main challenge is to adapt the measurement cycle frequency to the continuously harvested power.

As a consequence, Energy Harvesting can become a viable supply source for Wireless Sensor Networks of the future.

This chapter focuses on Vibration Energy Harvesting that can become an interesting power source for WSN in industrial environments with low light or no light at all. We will specifically concentrate on electrostatic devices, based on capacitive architectures, that are not as well-known as piezoelectric or electromagnetic devices, but that can present many advantages compared to them.

The next paragraph introduces the general concept of Vibration Energy Harvesters (VEH) and of electrostatic devices.

## 2. Vibration Energy Harvesting & Electrostatic Devices

Vibration Energy Harvesting is a concept that began to take off in the 2000's with the growth of MEMS devices. Since then, this concept has spread and conquered macroscopic devices as well.

### 2.1. Vibration Energy Harvesters – Overview

The concept of Vibration Energy Harvesting is to convert vibrations in an electrical power. Actually, turning ambient vibrations into electricity is a two steps conversion (Figure 5(a)). Vibrations are firstly converted in a relative motion between two elements, thanks to a mass-spring system, that is then converted into electricity thanks to a mechanical-to-electrical converter (piezoelectric material, magnet-coil, or variable capacitor). As ambient vibrations are generally low in amplitude, the use of a mass-spring system generates a phenomenon of resonance, amplifying the relative movement amplitude of the mobile mass compared to the vibrations amplitude, increasing the harvested power (Figure 5(b)).

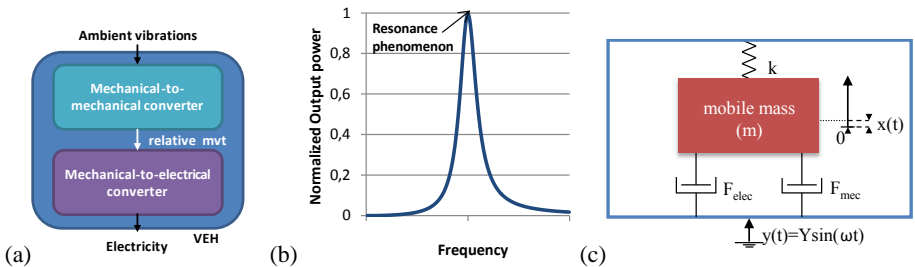

Figure 5. Vibration Energy Harvesters (a) concept (b) resonance phenomenon and (c) model

Figure 5(c) represents the equivalent model of Vibration Energy Harvesters. A mass (m) is suspended in a frame by a spring (k) and damped by forces ($f_{elec}$ and $f_{mec}$). When a vibration occurs $y(t) = Y \sin(\omega t)$, it induces a relative motion of the mobile mass $x(t) = X \sin(\omega t + \varphi)$ compared to the frame. A part of the kinetic energy of the moving mass is converted into electricity (modeled by an electromechanical force $f_{elec}$), while an other part is lost in friction forces (modeled by $f_{mec}$).

Newton's second law gives the differential equation that rules the moving mass's relative movement (equation 1). Generally, the mechanical friction force can be modeled as a viscous force $f_{mec} = b_m \dot{x}$.

Electrostatic Conversion for Vibration Energy Harvesting

Then, the equation of movement can be simplified by using the natural angular frequency $\omega_0 = \sqrt{k/m}$ and the mechanical quality factor $Q_m = m\omega_0/b_m$.

$$m\ddot{x} + f_{meca} + kx + f_{elec} = -m\ddot{y} \Rightarrow \ddot{x} + \frac{\omega_0}{Q_m}\dot{x} + \omega_0^2 x + \frac{f_{elec}}{m} = -\ddot{y} \qquad (1)$$

Then, when the electromechanical and the friction forces can be modeled by viscous forces, $f_{elec} = b_e \dot{x}$ and $f_{mec} = b_m \dot{x}$, where $b_e$ and $b_m$ are respectively electrical and mechanical damping coefficients, William and Yates [2] have proven that the maximum output power of a resonant energy harvester submitted to an ambient vibration is reached when the natural angular frequency ($\omega_0$) of the mass-spring system is equal to the angular frequency of ambient vibrations ($\omega$) and when the damping rate $\xi_e = b_e/(2m\omega_0)$ of the electrostatic force $f_{elec}$ is equal to the damping rate $\xi_m = b_m/(2m\omega_0)$ of the mechanical friction force $f_{mec}$. This maximum output power $P_{W\&Y}$ can be simply expressed with (2), when $\xi_e = \xi_m = \xi = 1/(2Q_m)$.

$$P_{W\&Y} = \frac{mY^2\omega_0^3 Q_m}{8} \qquad (2)$$

But obviously, to induce this electromechanical force, it is necessary to develop a mechanical-to-electrical converter to extract a part of mechanical energy from the mass and to turn it into electricity.

## 2.2. Converters & Electrostatic devices – Overview

Three main converters enable to turn mechanical energy into electricity: piezoelectric devices, electromagnetic devices and electrostatic devices (Table 1).

- Piezoelectric devices: they use piezoelectric materials that present the ability to generate charges when they are under stress/strain.
- Electromagnetic devices: they are based on electromagnetic induction and ruled by Lenz's law. An electromotive force is generated from a relative motion between a coil and a magnet.
- Electrostatic devices: they use a variable capacitor structure to generate charges from a relative motion between two plates.

Table 1. Mechanical-to-electrical converters for small-scale devices

| Piezoelectric converters | Electromagnetic converters | Electrostatic converters |
|---|---|---|
| Use of piezoelectric materials | Use of Lenz's law | Use of a variable capacitor structure |
| 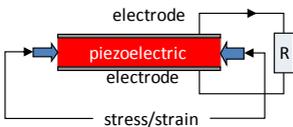 | 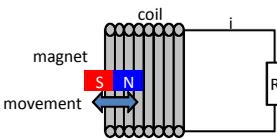 | 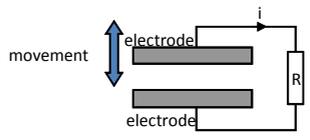 |

Obviously, each of these converters presents both advantages and drawbacks depending on the application (amplitudes of vibrations, frequencies…).

### 2.3. Advantages & Drawbacks of Electrostatic Devices

A summary of advantages and drawbacks of electrostatic devices is presented in Table 2. In most cases, piezoelectric and electrostatic devices are more appropriate for small scale energy harvesters (<1-10 cm³) while electromagnetic converters are better for larger devices.

Table 2. Advantages and drawbacks of converters

| | Piezoelectric devices | Electromagnetic devices | Electrostatic devices |
|---|---|---|---|
| Advantages | -high output voltages<br>-high capacitances<br>-no need to control any gap | -high output currents<br>-long lifetime proven<br>-robustness | -high output voltages<br>-possibility to build low-cost systems<br>-coupling coefficient easy to adjust<br>-high coupling coefficients reachable<br>-size reduction increases capacitances |
| Drawbacks | -expensive (material)<br>-coupling coefficient linked to material properties | -low output voltages<br>-hard to develop MEMS devices<br>-may be expensive (material)<br>-low efficiency in low frequencies and small sizes | -low capacitances<br>-high impact of parasitic capacitances<br>-need to control µm dimensions<br>-no direct mechanical-to-electrical conversion for electret-free converters |

This chapter is focused on electrostatic vibration energy harvesters. These VEH are well-adapted for size reduction, increasing electric fields, capacitances and therefore converters' power density capabilities. They also offer the possibility to decouple the mechanical structure and the converter (which is not possible with piezoelectric devices). Finally, they can be a solution to increase the market of EH-powered WSN by giving the possibility to develop "low-cost" devices as they do not need any magnet or any piezoelectric material that can be quite expensive.

The next paragraph is aimed at presenting the conversion principles of electrostatic devices. It covers both standard (electret-free) and electret-based electrostatic converters.

### 2.4. Conversion principles

Electrostatic converters are capacitive structures made of two plates separated by air, vacuum or any dielectric materials. A relative movement between the two plates generates a capacitance variation and then electric charges. These devices can be divided into two categories:

- Electret-free electrostatic converters that use conversion cycles made of charges and discharges of the capacitor (an active electronic circuit is then required to apply the charge cycle on the structure and must be synchronized with the capacitance variation).
- Electret-based electrostatic converters that use electrets, giving them the ability to directly convert mechanical power into electricity.

Electrostatic Conversion for Vibration Energy Harvesting

### 2.4.1. Electret-free electrostatic converters

These first electrostatic devices are passive structures that require an energy cycle to convert mechanical energy into electricity. Many energy cycles enable such a conversion, but the most commonly-used are charge-constrained and voltage-constrained cycles (Figure 6). They both start when the converter's capacitance is maximal. At this point, a charge is injected into the capacitor thanks to an external source, to polarize it. Charge-constrained and voltage-constrained cycles are presented in the following sub-sections.

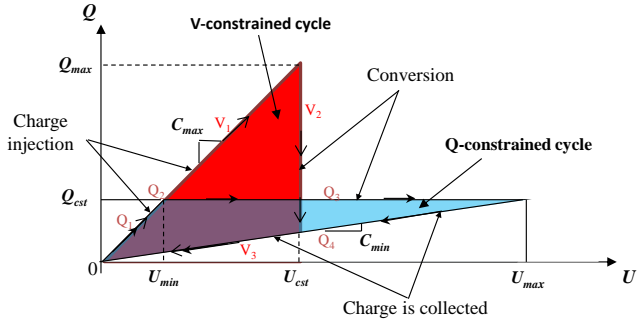

Figure 6. Standard energy conversion cycles for electret-free electrostatic devices

1.  **Charge-constrained Cycle**

The charge-constrained cycle (Figure 7) is the easiest one to implement on electrostatic devices. The cycle starts when the structure reaches its maximum capacitance $C_{max}$ ($Q_1$). In this position, the structure is charged thanks to an external polarization source: an electric charge $Q_{cst}$ is stored in the capacitor under a given voltage $U_{min}$. The device is then let in open circuit ($Q_2$). The structure moves mechanically to a position where its capacitance is minimal ($Q_3$). As the charge $Q_{cst}$ is kept constant while the capacitance $C$ decreases, the voltage across the capacitor $U$ increases. When the capacitance reaches its minimum ($C_{min}$) (or the voltage its maximum ($U_{max}$)), electric charges are removed from the structure ($Q_4$).

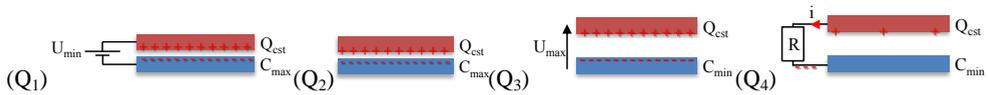

Figure 7. Charge-constrained cycle

The total amount of energy converted at each cycle is presented in equation (3).

$$E_{Q=cte} = \frac{1}{2}Q_{cst}^2 \left(\frac{1}{C_{min}} - \frac{1}{C_{max}}\right)$$  (3)

2.  **Voltage-Constrained Cycle**

The voltage-constrained cycle (Figure 8) also starts when the capacitance of the electrostatic converter is maximal. The capacitor is polarized at a voltage $U_{cst}$ using an external supply source (battery, charged capacitor…) ($V_1$). This voltage will be maintained throughout the conversion cycle thanks to an electronic circuit. Since the voltage is constant and the capacitance decreases, the charge of the capacitor increases, generating a current that is scavenged and stored ($V_2$). When the capacitance reaches its minimum value, the charge Q still presents in the capacitor is completely collected and stored ($V_3$).

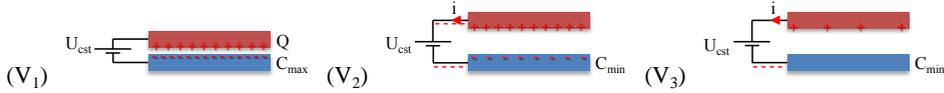

Figure 8. Voltage-constrained cycle

The total amount of energy converted at each cycle is presented in equation (4).

$$E_{U=cte} = U_{cst}^2 (C_{max} - C_{min}) \qquad (4)$$

In order to maximize the electrostatics structures' efficiency, a high voltage polarization source (>100V) is required. Obviously, this is a major drawback of these devices as it implies that an external supply source (battery, charged capacitor) is required to polarize the capacitor at the beginning of the cycle or at least at the first cycle (as one part of the energy harvested at the end of a cycle can be reinjected into the capacitor to start the next cycle).

One solution to this issue consists in using electrets, electrically charged dielectrics, that are able to polarize electrostatic energy harvesters throughout their lives, avoiding energy cycles and enabling a direct mechanical-to-electrical conversion. Electrostatic energy harvesters developed today tend to use them increasingly.

### 2.4.2.   Electret-Based Electrostatic converter

Electret-based electrostatic converters are quite similar to electret-free electrostatic converters. The main difference relies on the electret layers that are added on one (or two) plate(s) of the variable capacitor, polarizing it.

1.   Electrets

Electrets are dielectric materials that are in a quasi-permanent electric polarization state (electric charges or dipole polarization). They are electrostatic dipoles, equivalent to permanent magnets (but in electrostatic) that can keep charges for years. The word **electret** comes from "**electr**icity magn**et**" and was chosen by Oliver Heaviside in 1885.

a.   Definition and electret types

Electret's polarization can be obtained by dipole orientation (Figure 9(a)) or by charge injection (Figure 9(b)) leading to two different categories of electrets:

- Oriented-dipole electrets (dipole orientation)
- Real-charge electrets (excess of positive or negative charges on the electret's surface or on the electret's volume)

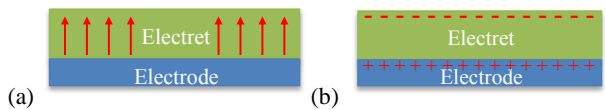

Figure 9. Standard electrets for electret-based electrostatic converters (a) dipole orientation and (b) charge injection

Electrostatic Conversion for Vibration Energy Harvesting

In the past, electrets were essentially obtained thanks to dipole orientation, from Carnauba wax for example [3]. Today, real-charge electrets are the most commonly used and especially in vibration energy harvesters because they are easy to manufacture with standard processes.

Indeed, oriented-dipole electrets and real-charge electrets are obtained from very different processes, leading to different behaviors.

b. Fabrication Processes

The first step to make oriented-dipole electrets is a heating of a dielectric layer above its melting temperature. Then, an electric field is maintained throughout the dielectric layer when it is cooling down. This enables to orient dielectric layer's dipoles in the electric field's direction. Solidification enables to keep dipoles in their position. This manufacturing process is similar to the one of magnets.

As for real-charge electrets, they are obtained by injecting an excess of charges in a dielectric layer. Various processes can be used: electron beam, corona discharge, ion or electron guns.

Here, we focus on charge injection by a triode corona discharge, which is probably the quickest way to charge dielectrics. Corona discharge (Figure 10) consists in a point-grid-plane structure whose point is submitted to a strong electric field: this leads to the creation of a plasma, made of ions that are projected onto the surface of the sample to charge and whose charges are transferred to the dielectric layer's surface.

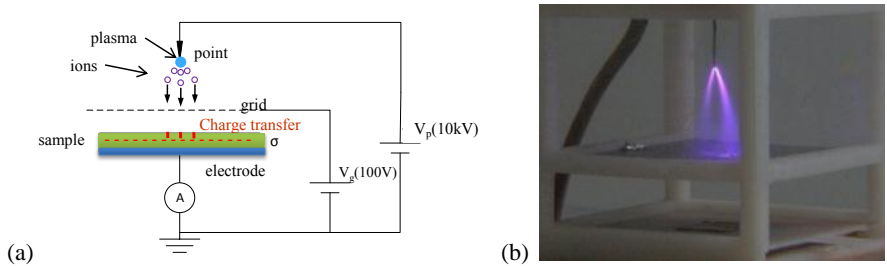

Figure 10. Corona discharge device (a) principle and (b) photo (CEA-LETI)

Corona discharge may be positive or negative according to the sign of the point's voltage. Positive and negative corona discharges have different behaviors as the plasma and the charges generated are different. Obviously positive corona discharges will lead to positively-charged electrets and negative corona discharges to negatively-charged electrets that have also different behaviors (stability, position of the charges in the dielectric). The grid is used to control the electret's surface voltage $V_s$ that results from the charges injected. Actually, when the electret's surface voltage $V_s$ reaches the grid voltage $V_g$, there is no potential difference between the grid and the sample any longer and therefore, no charge circulation anymore. So, at the end of the corona charging, the electret's surface voltage is equal to the grid voltage.

c. Equivalent model of electrets

Charge injection or dipole orientation leads to a surface potential $V_s$ on the electret (Gauss's law). It is generally assumed that charges are concentrated on the electret's surface and therefore, the surface potential can be simply expressed by: $V_s = \sigma d / \varepsilon \varepsilon_0$, with ε the electret's dielectric permittivity, σ its surface charge density and d its thickness.

The equivalent model of an electret layer (Figure 11(a) and (b)) is then a capacitor $C = \varepsilon \varepsilon_0 S / d$ in series with a voltage source whose value is equal to the surface voltage of the electret $V_s$ (Figure 11(c)).

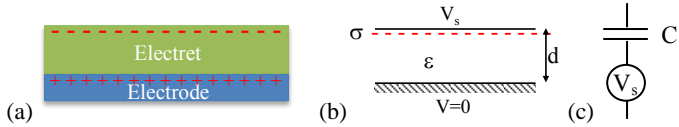

Figure 11. (a) electret layer (b) parameters and (c) equivalent model

### d. Charge stability and measurement

Nevertheless and unfortunately, dielectrics are not perfect insulators. As a consequence, some charge conduction phenomena may appear in electrets, and implanted charges can move inside the material or can be compensated by other charges or environmental conditions, and finally disappear. Charge stability is a key parameter for electrets as the electret-based converter's lifetime is directly linked to the one of the electret. Therefore, it is primordial to choose stable electrets to develop electret-based vibration energy harvesters.

Many measurement methods have been developed to determine the quantity of charges stored into electrets and their positions. These methods are really interesting to understand what happens in the material but are complicated to implement and to exploit. Yet, for vibration energy harvesters, the most important data is the surface potential decay (SPD), that is to say, the electret's surface voltage as a function of the time after charging.

In fact, the surface voltage can be easily measured thanks to an electrostatic voltmeter (Figure 12(a)). This method is really interesting as it enables to make the measurement of the surface voltage without any contact and therefore without interfering with the charges injected into the electret.

Figure 12(b) presents some examples of electrets' SPDs (good, fair and poor stability). Electret stability depends of course of the dielectric material used to make the electret: dielectrics that have high losses (high $\tan(\delta)$) are not good electrets and may lose their charges in some minutes, while materials such as Teflon or silicon dioxide ($SiO_2$) are known as stable electrets. But, other parameters, such as the initial surface voltage or the environmental conditions (temperature, humidity) have an important impact as well. Generally, for a given material, the higher the initial surface voltage is, the lower the stability becomes. For environmental conditions, high temperatures and high humidity tend to damage the electret stability.

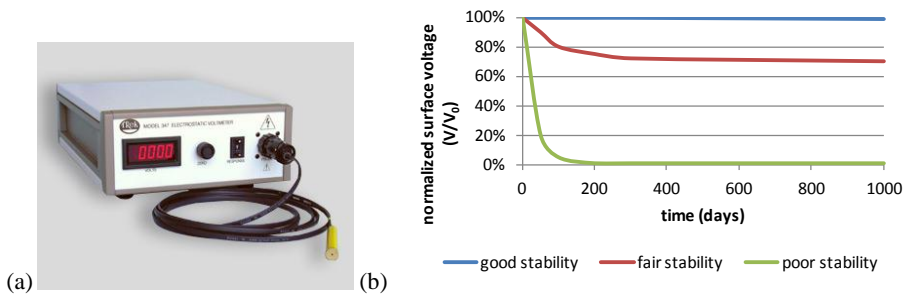

Figure 12. (a) Electrostatic voltmeter (Trek® 347) and (b) examples of Surface Potential Decays (SPDs)

Electrostatic Conversion for Vibration Energy Harvesting

The electret behaviors of many materials have been tested. The next sub-section gives some examples of well-known and stable electrets.

e. Well-known electrets

Teflon [4-7], SiO$_2$ [8-9] and CYTOP [10-14] are clearly the most well-known and the most used electrets in electret-based electrostatic converters. Of course, many more electrets can be found in the state of the art. Table 3 presents some properties of these electrets. It is for example interesting to note that SiO$_2$-based electrets have the highest surface charge densities. As for the stability, it is quite complicated to provide a value. Actually, it greatly depends on the storage, the humidity, the temperature, the initial conditions, the thickness… Yet, the examples given below show a stability V$_{s90\%}$ (90% of the initial surface voltage) generally higher than 2-3 years.

Table 3. Well-known electrets from the state of the art

| Electret | Deposition method | Maximum thickness | Dielectric Strength (V/µm) | ε | Standard surface charge density (mC/m²) |
|---|---|---|---|---|---|
| Teflon (PTFE/FEP/PFA) | Films are glued | Some 100 µm | 100-140 | 2.1 | 0.1-0.25 |
| SiO$_2$-based electrets | Thermal oxidization of silicon wafers (+ LPCVD Si$_3$N$_4$) | Some µm (<3µm) | 500 | 4 | 5-10 |
| Parylene (C/HT) | PVD-like deposition method | Some 10µm | 270 | 3 | 0.5-1 |
| CYTOP | Spin-coating | 20µm | 110 | 2 | 1-2 |
| Teflon AF | Spin-coating | 20µm | 200 | 1.9 | 0.1-0.25 |

Added in capacitive structures, electrets enable a simple mechanical-to-electrical energy conversion.

2. Conversion principle

The conversion principle of electret-based electrostatic converters is quite similar to electret-free electrostatic converters and is tightly linked to variations of capacitance. But contrary to them, the electret-based conversion does not need any initial electrical energy to work; a structure deformation induces directly an output voltage, just like a piezoelectric material.

a. Principle

Electret-based converters are electrostatic converters, and are therefore based on a capacitive structure made of two plates (electrode and counter-electrode (Figure 13)). The electret induces charges on electrodes and counter-electrodes to respect Gauss's law. Therefore, Q$_i$, the charge on the electret is equal to the sum of Q$_1$ and Q$_2$, where Q$_1$ is the total amount of charges on the electrode and Q$_2$ the total amount of charges on the counter-electrode (Q$_i$=Q$_1$+Q$_2$). A relative movement of the counter-electrode compared to the electret and the electrode induces a change in the capacitor geometry (e.g. the counter-electrode moves away from the electret, changing the air gap and then the electret's influence on the counter-electrode) and leads to a reorganization of charges between the electrode and the counter-electrode through load R (Figure 14). This results in a current circulation through R and one part of the mechanical energy (relative movement) is then turned into electricity.

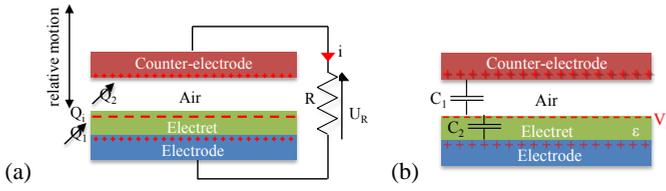

Figure 13. Electret-based electrostatic conversion – Concept

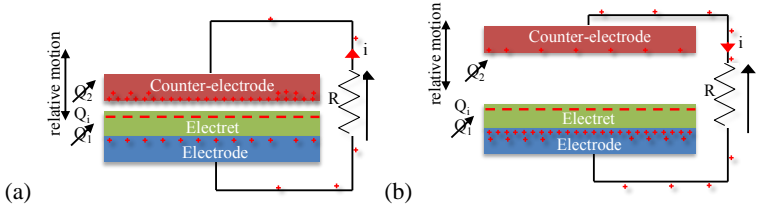

Figure 14. Electret-based electrostatic conversion – Charge circulation

The equivalent model of electret-based electrostatic converters is presented below.

b.    Equivalent model and Equations

The equivalent model of electret-based electrostatic converters is quite simple as it consists in a voltage source in series with a variable capacitor. This model has been confronted to experimental data and corresponds perfectly to experimental results (see section 3.2.4).

Figure 15 presents an electret-based electrostatic converter connected to a resistive load R. As the capacitances of electret-based converters are quite low (often lower than 100pF), it is important to take parasitic capacitances into account. They can be modeled by a capacitor in parallel with the electret-based converter. And actually, only 10pF of parasitic capacitances may have a deep impact on the electret-based converter's output voltages and output powers.

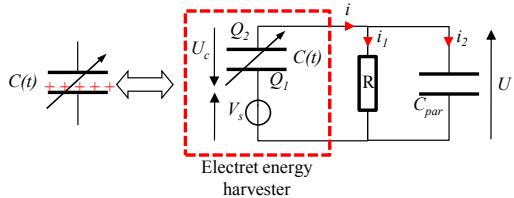

Figure 15. Electrical equivalent model of electret-based electrostatic converters

In the case of a simple resistive load placed at the terminals of the electret-based converter, the differential equation that rules the system is presented in equation (5).

$$\frac{dQ_2}{dt} = \frac{V_s}{R} - \frac{Q_2}{C(t)R} \tag{5}$$

Taking parasitic capacitances into account, this model is modified into (6) [15].

Electrostatic Conversion for Vibration Energy Harvesting

$$\frac{dQ_2}{dt} = \frac{1}{\left(1 + \frac{C_{par}}{C(t)}\right)} \left( \frac{V_s}{R} - Q_2 \left( \frac{1}{RC(t)} - \frac{C_{par}}{C(t)^2} \frac{dC(t)}{dt} \right) \right) \qquad (6)$$

Actually, electret-based converter's output powers (P) are directly linked to the electret's surface voltage $V_s$ and the capacitance variation $dC/dt$ when submitted to vibrations [16]. And, as a first approximation:

$$P \propto V_s^2 \frac{dC}{dt} \qquad (7)$$

As a consequence, as the electret-based converter's output powers is linked to the electret's surface voltage $V_s$ and its lifetime to the electret's lifetime, we confirm that Surface Potential Decays (SPDs) are the most appropriate way to characterize electrets for an application in energy harvesting.

### 2.4.3. Capacitors and capacitances' models

Whether it is electret-free or electret-based conversion, electrostatic converters are based on a variable capacitive structure. This subsection is focused on the main capacitor shapes employed in electrostatic converters and on their models.

1. Main capacitor shapes

Most of the electrostatic converters' shapes are derived from accelerometers. Actually, it is possible to count four main capacitor shapes for electrostatic converters (Figure 16).

(a) in-plane gap closing converter: interdigitated comb structure with a variable air gap between fingers and movement in the plane

(b) in-plane overlap converter: interdigitated comb structure with a variable overlap of the fingers and movement in the plane

(c) out-of-plane gap closing converter: planar structure with a variable air gap between plates and perpendicular movement to the plane

(d) in-plane converter with variable surface: planar structure with a variable overlap of the plates and movement in the plane. There is a great interest of developing patterned versions with bumps and trenches facing.

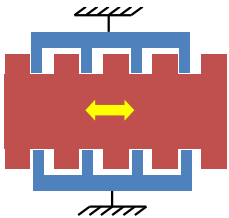
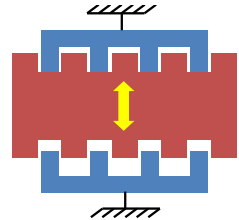

(a) in-plane gap closing converter          (b) in-plane overlap converter

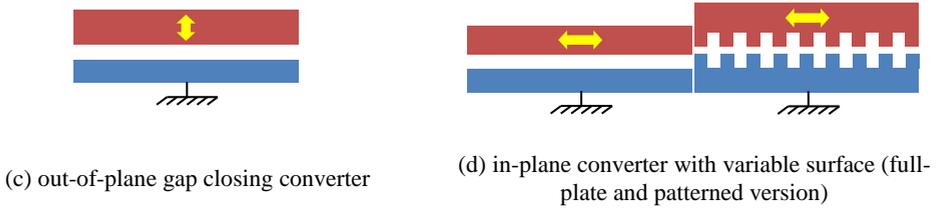

(c) out-of-plane gap closing converter     (d) in-plane converter with variable surface (full-plate and patterned version)

Figure 16. Basic capacitor shapes for electrostatic converters

Obviously, these basic shapes can be adapted to electret-free and electret-based electrostatic converters. As capacitances and electrostatic forces are heavily dependent on the capacitor's shape and on its dimensions, it is interesting to know the capacitance and the electrostatic forces generated by each of these structures to design an electrostatic converter. Capacitances values and electrostatic forces for each shape are presented in the next sub-section.

### 2. Capacitances values and electrostatic forces

Capacitances values are all deduced from the simple plane capacitor model. In this subsection, the capacitance is computed with an electret layer. To get the capacitances for electret-free electrostatic converters, one has just to take d=0 (where d is the electret thickness).

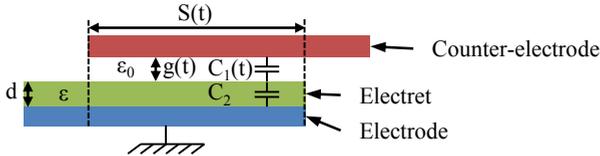

Figure 17. Capacitance of the simple plane capacitor

The total capacitance of the electrostatic converter presented in Figure 17 corresponds to two capacitances ($C_1$ and $C_2$) in series.

$$C(t) = \frac{C_1(t)C_2}{C_1(t)+C_2} = \frac{\varepsilon_0 S(t)}{g(t) + d/\varepsilon} \qquad (8)$$

The electrostatic force $f_{elec}$ induced by this capacitor can be expressed by:

$$F_{elec} = \frac{d}{dx}\big(W_{elec}\big) = \frac{d}{dx}\left(\frac{1}{2}C(x)U_c(x)^2\right) = \frac{d}{dx}\left(\frac{1}{2}\frac{Q_c^2(x)}{C(x)}\right) \qquad (9)$$

With $W_{elec}$ the total amount of electrostatic energy stored in C, $Q_c$ the charge on C, $U_c(x)$ the voltage across C and x the relative movement of the upper plate compared to the lower plate.

Capacitances and electrostatic forces of the four capacitor shapes are obtained by integrating equations (8) and (9).

### a. In-plane gap-closing converter

Electrostatic Conversion for Vibration Energy Harvesting

In-plane gap-closing converters are interdigitated comb devices with a variable air gap between fingers as presented in Figure 18.

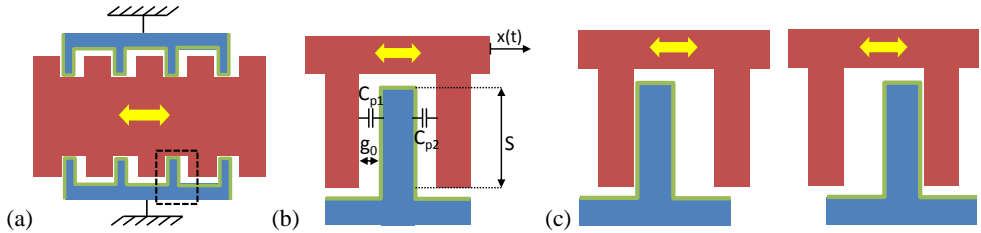

Figure 18. (a) In-plane gap-closing converters, (b) zoom on one finger with $C_{min}$ position, (c) $C_{max}$ positions

The capacitance of the converter corresponds to the two capacitors $C_{p1}$ and $C_{p2}$ in parallel and is expressed in equation (10).

$$C(x) = \frac{2N\varepsilon_0 S\left(g_0 + \frac{d}{\varepsilon}\right)}{\left(g_0 + \frac{d}{\varepsilon}\right)^2 - x^2} \qquad (10)$$

Where N is the number of fingers of the whole electrostatic converter, and S the facing surface.

b. In-plane overlap converter

In-plane overlap converters are interdigitated comb structure with a variable overlap of the fingers as presented in Figure 19. The whole structure must be separated into two variable capacitors $C_{c1}$ and $C_{c2}$ as the increase of $C_{c1}$'s capacitance leads to a decrease of $C_{c2}$'s capacitance and vice-versa.

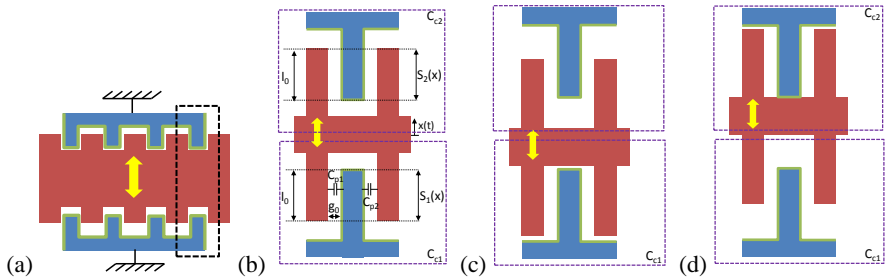

Figure 19. (a) In-plane overlap converters, (b) zoom on two fingers (c) $C_{max}$ position for $C_{c1}$ and $C_{min}$ position for $C_{c2}$ and (d) $C_{min}$ position for $C_{c1}$ and $C_{max}$ position for $C_{c2}$.

The capacitance of $C_{c1}$ and $C_{c2}$ are expressed in equation (11).

$$C_{c1}(x) = \frac{\varepsilon_0 N w}{g_0 + \frac{d}{\varepsilon}}\left(l_0 - x\right) \text{ and } C_{c2}(x) = \frac{\varepsilon_0 N w}{g_0 + \frac{d}{\varepsilon}}\left(l_0 + x\right) \qquad (11)$$

Where N is the number of fingers, $l_0$ the facing length of $C_{c1}$ and $C_{c2}$ at the equilibrium position and w the thickness of the fingers (third dimension).

c.   Out-of-plane gap closing converter

In this configuration, the counter-electrode moves above the electrode, inducing a variation of the gap between the electret and the counter-electrode. The initial gap between the counter-electrode and the electret is $g_0$ and the surface is denoted by S (Figure 20(a)). $C_{max}$ and $C_{min}$ positions are presented in Figure 20(b, c).

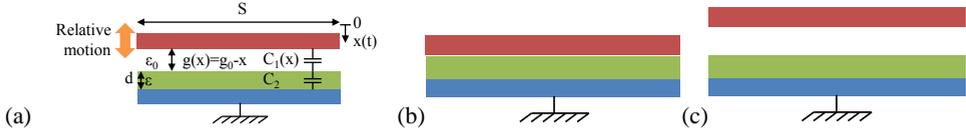

Figure 20. (a) Out-of-plane gap closing, (b) $C_{max}$ position and (c) $C_{min}$ position

The capacitance of the converter is expressed in equation (12).

$$C(x) = \frac{\varepsilon_0 S}{g_0 + d/\varepsilon - x} \qquad (12)$$

d.   In-plane converter with variable surface

In in-plane converters with variable surface, it is a change in the capacitor's area that is exploited, as presented in Figure 21(a).

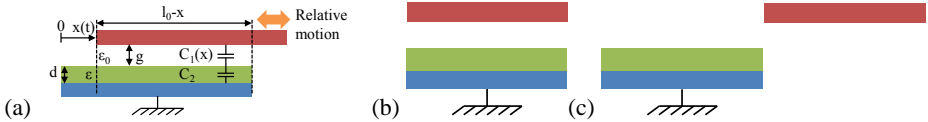

Figure 21. (a) In-plane with variable surface, (b) $C_{max}$ position and (c) $C_{min}$ position

The capacitance of the converter is expressed in equation (13).

$$C(x) = \frac{\varepsilon_0 w(l_0 - x)}{g + d/\varepsilon} \qquad (13)$$

Where w is the converter's thickness (third dimension), $l_0$ the facing length between the plates at t=0.

So as to increase the capacitance variation for a given relative displacement x, it is interesting to pattern the capacitive structure, as presented in Figure 22.

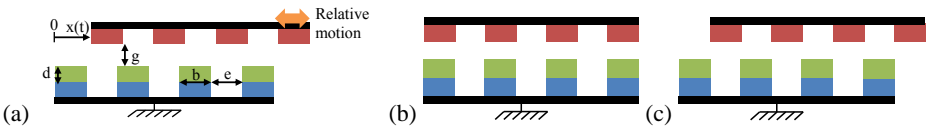

Figure 22. (a) In-plane converter with variable patterned surface, (b) $C_{max}$ position and (c) $C_{min}$ position

To develop efficient vibration energy harvesters able to work with low vibrations, it is necessary to use micro-patterned capacitive structure (small e and b). With such dimensions, fringe effects must be

Electrostatic Conversion for Vibration Energy Harvesting

taken into account and it was proven [17] that the capacitance of the energy harvester can be simply modeled by a sine function presented in equation (14).

$$C(x) = \frac{C_{max} + C_{min}}{2} + \left(\frac{C_{max} - C_{min}}{2}\right) \times \cos\left(\frac{2\pi x}{e+b}\right) \qquad (14)$$

Where $C_{max}$ and $C_{min}$ are the maximal and the minimal capacitances of the energy harvester and computed by finite elements.

Electrostatic forces are deduced from the derivation of the electrostatic energy stored into the capacitor, as presented in equation (9). Concerning electret-based devices, the electrostatic force cannot be easily expressed as both capacitor's charge and voltage change when the geometry varies. Table 4 overviews the electrostatic forces for the various converters and their operation modes.

Table 4. Electrostatic forces according to the converter and its operation mode

| Converter | $f_{elec}$ for charge-constrained cycle | $f_{elec}$ for voltage-constrained cycle | $f_{elec}$ for electret-based devices |
|---|---|---|---|
| In-plane gap closing | $\dfrac{Q_{cst}^2 x}{2\varepsilon_0 N\left(g_0 + d/\varepsilon\right)S}$ | $\dfrac{2N\varepsilon_0\left(g_0 + d/\varepsilon\right)SxU_{cst}^2}{\left(\left(g_0 + d/\varepsilon\right)^2 - x^2\right)^2}$ | $\dfrac{d}{dx}\left(W_{elec}\right)$ |
| In-plane overlap (only $C_{c2}$) | $\dfrac{Q_{cst}^2\left(g_0 + d/\varepsilon\right)}{2\varepsilon_0 Nw\left(l_0 + x\right)^2}$ | $\dfrac{\varepsilon_0 NwU_{cst}^2}{2\left(g_0 + d/\varepsilon\right)}$ | $\dfrac{d}{dx}\left(W_{elec}\right)$ |
| Out-of-plane gap closing | $\dfrac{Q_{cst}^2}{2\varepsilon_0 S}$ | $\dfrac{\varepsilon_0 SU_{cst}^2}{2\left(g_0 + d/\varepsilon - x\right)^2}$ | $\dfrac{d}{dx}\left(W_{elec}\right)$ |
| In-plane with variable surface (non patterned) | $\dfrac{Q_{cst}^2\left(g_0 + d/\varepsilon\right)}{2\varepsilon_0 w\left(l_0 - x\right)^2}$ | $\dfrac{\varepsilon_0 wU_{cst}^2}{2\left(g_0 + d/\varepsilon\right)}$ | $\dfrac{d}{dx}\left(W_{elec}\right)$ |

These electrostatic converters are then coupled to mass-spring systems to become vibration energy harvesters.

## 3. Electrostatic Vibration Energy Harvesters (eVEH)

As presented in section 2.1, harvesting vibrations requires two conversion steps: a mechanical-to-mechanical converter made of a mass-spring resonator that turns ambient vibrations into a relative movement between two elements (presented in 2.1) and a mechanical-to-electrical converter using, in our case, a capacitive architecture (presented in 2.2) that converts this relative movement into electricity. Section 3 is aimed at presenting complete devices that gather these two converters. It is firstly focused on electret-free electrostatic devices before presenting electret-based devices.

### 3.1. Electret-free Electrostatic Vibration Energy Harvesters (eVEH)

#### 3.1.1. Devices

The first MEMS electrostatic comb based VEH was developed at the MIT by Meninger et al. in 2001 [18]. This device used an in-plane overlap electrostatic converter. Operating cycles are described and it is proven that the voltage-constrained cycle enables to maximize output power (if the power management electronic is limited in voltage). Yet, for the prototype, a charge-constrained cycle was adopted to simplify the power management circuit even if it drives to a lower output power.

Electrostatic devices can be particularly suitable for Vibration energy harvesting at low frequencies (<100Hz). In 2002, Tashiro et al. [19] developed a pacemaker capable of harvesting power from heartbeats. The output power of this prototype installed on the heart of a goat was 58μW.

In 2003, Roundy [20] proved that the best structure for electrostatic devices was the in-plane gap closing and would be able to harvest up to 100μW/cm³ with ambient vibrations (2.25m/s²@120Hz). Roundy et al. then developed an in-plane gap closing structure able to harvest 1.4nJ/cycle.

In 2005, Despesse et al. developed a macroscopic device (Figure 23(a)) able to work on low vibration frequencies and able to harvest 1mW for a vibration of 0.2G@50Hz [21]. This prototype has the highest power density of eVEH ever reached. Some other MEMS devices were then developed by Basset et al [22] (Figure 23(b)) and Hoffmann et al. [23].

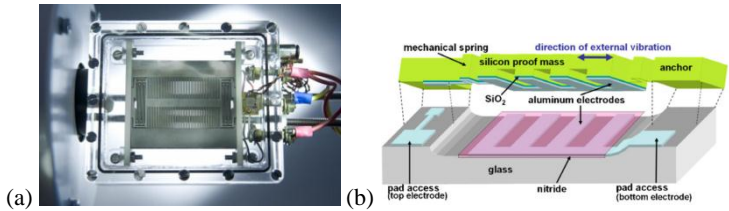

Figure 23. Electrostatic vibration energy harvesters from (a) Despesse et al. [21] and (b) Basset et al. [22].

#### 3.1.2. State of the art – Overview

An overview of electret-free electrostatic vibration energy harvesters is presented in Table 5.

Table 5. Electret-free electrostatic vibration energy harvesters from the state of the art

| Author | Ref | Output power | Surface | Volume | Polarization voltage | Vibrations |
|---|---|---|---|---|---|---|
| Tashiro | [19] | 36 μW | | 15000 mm³ | 45V | 1,2G@6Hz |
| Roundy | [24] | 11 μW | 100 mm² | 100 mm³ | | 0.23G@100Hz |
| Mitcheson | [25] | 24 μW | 784 mm² | 1568 mm³ | 2300 V | 0.4G@10Hz |
| Yen | [26] | 1,8 μW | 4356 mm² | 21780 mm³ | 6 V | 1560Hz |
| Despesse | [21] | 1050 μW | 1800 mm² | 18000 mm³ | 3 V | 0.3G@50Hz |
| Hoffmann | [23] | 3.5 μW | 30 mm² | | 50 V | 13G@1300-1500Hz |
| Basset | [22] | 61nW[1] | 66 mm² | 61.49mm³ | 8 V | 0.25G@250Hz |

[1] latest results from ESIEE showed that higher output powers are reachable thanks to this device (up to 500nW).

Electrostatic Conversion for Vibration Energy Harvesting

Many prototypes of electret-free electrostatic vibration energy harvesters have been developed and validated. Currently, the tendency is to couple these devices to electrets. The next subsection is focused on them.

### 3.2. Electret-Based Electrostatic VEH

Electret-based devices were developed to enable a direct vibration-to-electricity conversion (without cycles of charges and discharges) and to simplify the power management circuits.

#### 3.2.1. History

The idea of using electrets in electrostatic devices to make generators goes back to about 40 years ago. In fact, the first functional electret-based generator was developed in 1978 by Jefimenko and Walker [27]. From that time, several generators exploiting a mechanical energy of rotation were developed (Jefimenko [27], Tada [28], Genda [29] or Boland [16]). Figure 24 presents an example, developed by Boland in 2003 [16, 30] of an electret-based generator able to turn a relative rotation of the upper plate compared to the lower plate into electricity.

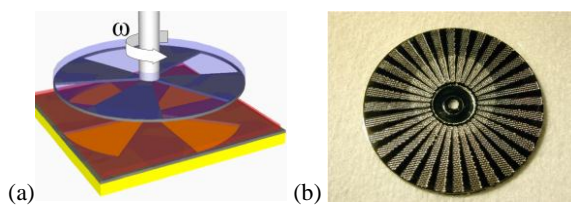

Figure 24. Boland's electret-based generator prototype [30] (a) perspective view and (b) stator

With the development of energy harvesting and the need to design autonomous sensors for industry, researchers and engineers have decided to exploit electrets in their electrostatic vibration energy harvesters as their everlasting polarization source.

#### 3.2.2. Devices

Even if the four capacitor shapes presented in subsection 2.3.1 are suitable to develop electret-based vibration energy harvesters, only two architectures have been really exploited: out-of-plane gap closing and patterned in-plane with variable surface structures.

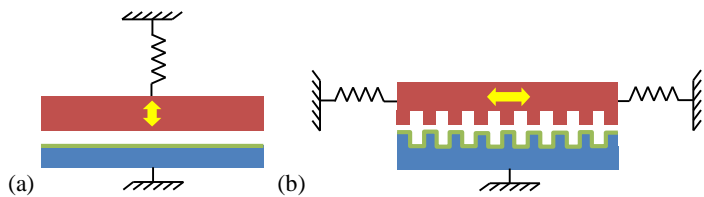

Figure 25. Standard architectures for electret-based Vibration Energy Harvesters

This section presents some examples of electret-based vibration energy harvesters from the state-of-the-art. We have decided to gather these prototypes in 2 categories: devices using full-sheet electrets (electret dimensions or patterning higher than 5mm) and devices using patterned electrets (electret dimensions or patterning smaller than 5mm). Indeed, it is noteworthy that texturing an electret is not an easy task as it generally leads to a weak stability (important charge decay) and requires MEMS fabrication facilities.

a.   Devices using full-sheet electrets

Full-sheet-electret devices can exploit a surface variation or a gap variation. In 2003, Mizuno [31] developed an out-of-plane gap closing structure using a clamped-free beam moving above an electret. This structure was also studied by Boisseau et al. [15] in 2011. This simple structure is sufficient to rapidly demonstrate the principle of vibration energy harvesting with electrets. Large amount of power can be harvested even with low vibration levels as soon as the resonant frequency of the harvester is tuned to the frequency of ambient vibrations.

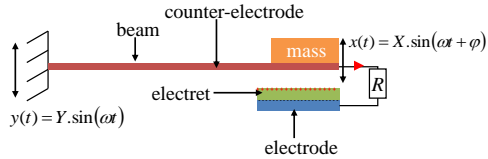

Figure 26. Cantilever-based electret energy harvesters [15]

The first integrated structure using full-sheet electrets was developed by Sterken et al. from IMEC [32] in 2007. A diagram is presented in Figure 27: a full-sheet is used as the polarization source. The electret layer polarizes the moving electrode of the variable capacitance ($C_{var}$). The main drawback of this prototype is to add a parasitic capacitance in series with the energy harvester, limiting the capacitance's variation and the converter's efficiency.

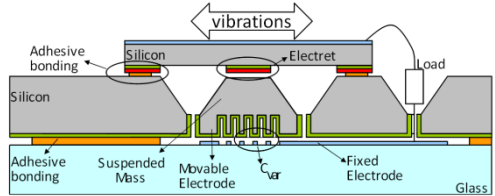

Figure 27. IMEC's first electret-based vibration energy harvester [32]

Today, most of the electret-based vibration energy harvesters use patterned electrets and exploit surface variation.

b.   Devices using patterned electrets

The first structure using patterned electrets was developed by the university of Tokyo in 2006 [33]. Many other devices followed, each of them, improving the first architecture [10, 34-38]. For example Miki et al. [39] improved these devices by developing a multiphase system and using non-linear effects. Multiphase devices enable to limit the peaks of the electrostatic force and thus to avoid to block the moving mass.

Electrostatic Conversion for Vibration Energy Harvesting

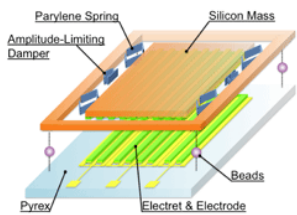

Figure 28. Multiphase electret energy harvester exploiting non-linear springs [38-39].

c. Mechanical springs to harvest ambient vibrations

Developing low-resonant frequency energy harvesters is a big challenge for small-scale devices. In most cases, ambient vibrations' frequencies are below 100Hz. This leads to long and thin springs difficult to obtain by using silicon technologies (form factors are large and structures become brittle). Thus, to reduce the resonant frequency of vibration energy harvesters, keeping small dimensions, solutions such as parylene springs [40] were developed. Another way consists in using microballs that act like a slideway. Naruse has already shown that such a system could operate at very low frequencies (<2 Hz) and could produce up to 40 μW [37] (Figure 29).

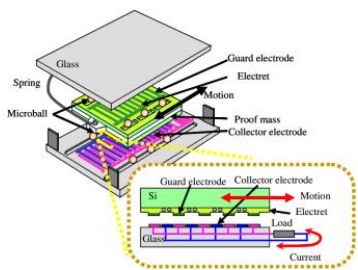

Figure 29. Device on microballs from [37]

Besides, a good review on MEMS electret energy harvesters can be found in [41]. The next subsection presents an overview of some electret-based prototypes from the state of the art.

### 3.2.3. State of the art – Overview

An overview of electret-based electrostatic vibration energy harvesters is presented in Table 6.

Table 6. Electret-based energy harvesters from the state of the art

| Author | Ref | Vibrations / Rotations | Active Surface | Electret Potential | Output Power |
|---|---|---|---|---|---|
| Jefimenko | [27] | 6000 rpm | 730 cm² | 500V | 25 mW |
| Tada | [28] | 5000 rpm | 90 cm² | 363V | 1.02 mW |
| Boland | [16] | 4170 rpm | 0.8 cm² | 150V | 25 μW |
| Genda | [29] | 1'000'000 rpm | 1.13 cm² | 200V | 30.4 W |
| Boland | [42] | 7.1G@60Hz | 0.12 cm² | 850V | 6 μW |
| Tsutsumino | [33] | 1.58G@20Hz | 4 cm² | 1100V | 38 μW |
| Lo | [43] | 14.2G@60Hz | 4.84 cm² | 300V | 2.26 μW |
| Sterken | [32] | 1G@500Hz | 0.09 cm² | 10V | 2nW |

| Lo | [34] | 4.93G@50Hz | 6 cm² | 1500V | 17.98 µW |
| Zhang | [35] | 0.32G@9Hz | 4 cm² | 100V | 0.13 pW |
| Yang | [44] | 3G@560Hz | 0.3 cm² | 400V | 46.14 pW |
| Suzuki | [40] | 5.4G@37Hz | 2.33 cm² | 450V | 0.28 µW |
| Sakane | [10] | 0.94G@20Hz | 4 cm² | 640V | 0.7 mW |
| Naruse | [37] | 0.4G@2Hz | 9 cm² | | 40µW |
| Halvorsen | [45] | 3.92G@596Hz | 0.48 cm² | | 1µW |
| Kloub | [46] | 0.96G@1740Hz | 0.42 cm² | 25V | 5µW |
| Edamoto | [36] | 0.87G@21Hz | 3 cm² | 600 V | 12µW |
| Miki | [39] | 1.57G@63Hz | 3 cm² | 180V | 1µW |
| Honzumi | [47] | 9.2G@500Hz | 0.01 cm² | 52V | 90 pW |
| Boisseau | [15] | 0.1G@50Hz | 4.16cm² | 1400V | 50µW |

Table 6 shows a significant increase of electret-based prototypes since 2003. It is also interesting to note that some companies such as Omron or Sanyo [48] started to study these devices and to manufacture some prototypes.

Thanks to simple cantilever-based devices developed for example by Mizuno [31] and Boisseau [15], the theoretical model of electret-based devices can be accurately validated.

### 3.2.4. Validation of theory with experimental data – Cantilever-based electret energy harvesters

The theoretical model of electret-based energy converters and vibration energy harvesters can be easily validated by experimental data with a simple cantilever-based electret vibration energy harvester [15].

a.  Device

The prototype presented in Figure 30 consists in a clamped-free beam moving with regards to an electret due to ambient vibrations. The mechanical-to-mechanical converter is the mass-beam system and the mechanical-to-electrical converter is made of the electrode-electret-airgap-moving counter-electrode architecture [15].

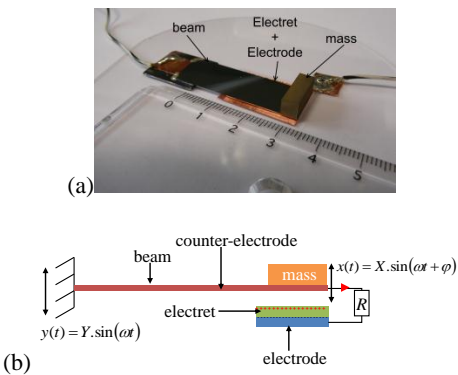

| $M_{beam}$ | Material of the beam | Silicon |
|---|---|---|
| $E$ | Young's Modulus of Silicon | 160 GPa |
| $L$ | Position of the centre of gravity of the mass | 30 mm |
| $h$ | Thickness of the beam | 300 µm |
| $w$ | Width of the beam / Width of the electret | 13 mm |
| $2Lm$ | Length of the mobile mass | 4 mm |
| $m$ | Mobile mass | 5 g |
| $\omega_v = \omega$ | Angular frequency of vibrations | 2π×50 rad/s |
| $Q_m$ | Mechanical quality factor of the structure | 75 |
| $M_{electret}$ | Material of the electret | FEP |
| $\varepsilon_\rho$ | Dielectric constant of the electret | 2 |

Electrostatic Conversion for Vibration Energy Harvesting

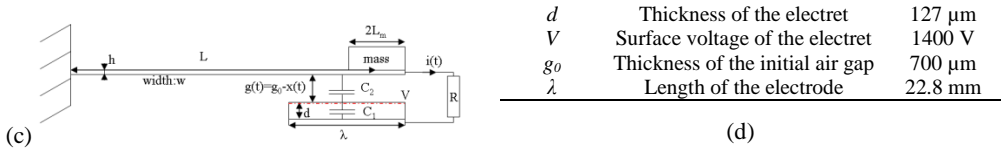

| | | |
|---|---|---|
| $d$ | Thickness of the electret | 127 µm |
| $V$ | Surface voltage of the electret | 1400 V |
| $g_0$ | Thickness of the initial air gap | 700 µm |
| $\lambda$ | Length of the electrode | 22.8 mm |

(c)                                                                 (d)

Figure 30. Example of a simple out-of-plane electret-based VEH (cantilever) (a) prototype, (b) diagram, (c) parameters and (d) dimensions

This system can be modeled by equations developed in section 2.

b.    Model

From equations (1) and (6), one can prove that this device is ruled by the system of differential equations (15).

$$
\begin{cases}
m\ddot{x} + b_m \cdot \dot{x} + kx - \dfrac{d}{dx}\left(\dfrac{Q_2^2}{2C(t)}\right) - mg = -m\ddot{y} \\[3mm]
\dfrac{dQ_2}{dt} = \dfrac{1}{\left(1 + \dfrac{C_{par}}{C(t)}\right)}\left(\dfrac{V}{R} - Q_2\left(\dfrac{1}{RC(t)} - \dfrac{C_{par}}{C(t)^2}\dfrac{dC(t)}{dt}\right)\right)
\end{cases}
\quad (15)
$$

Obviously, this system cannot be solved by hand. Yet, by using a numerical solver (e.g. Matlab), this becomes possible. It is also imaginable to use Spice by turning this system of equations in its equivalent electrical circuit (Figure 31).

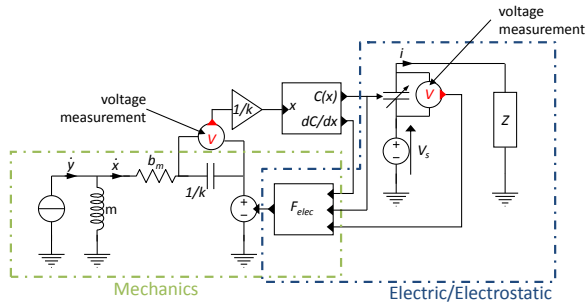

Figure 31. Equivalent electrical model of electret-based vibration energy harvesters

c.    Theory vs experimental data

The prototype presented in Figure 30 has been tested on a shaker at 0.1G@50Hz with two different loads (300MΩ and 2.2GΩ) and the corresponding theoretical results have been computed using a numerical solver. Theoretical and experimental output voltages are presented in Figure 32 showing an excellent match.

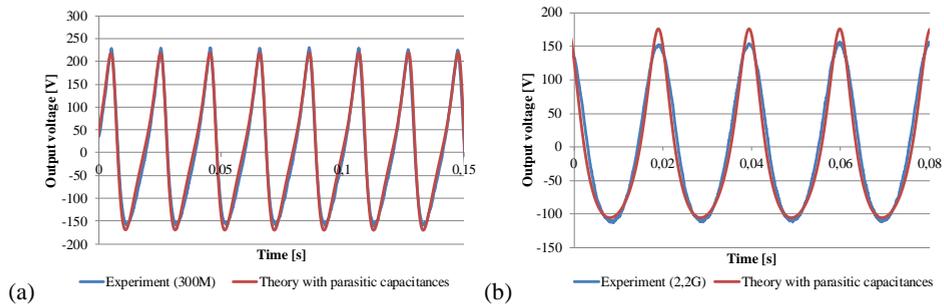

(a)                                                                (b)

Figure 32. Validation of theory with a cantilever-based electret energy harvester (a) R=300MΩ and (b) R=2.2GΩ

This simple prototype enables to validate the model of electret-based vibration energy harvesters that was presented in section 2. It is also interesting to note that this simple prototype has an excellent output power that reaches 50µW with a low vibration acceleration of 0.1G@50Hz.

Section 3 is concluded by an overview of electret patterning methods. Actually, electret patterning can be a real challenge in electret-based devices because of weak stability problems.

### 3.2.5. Electret patterning

As presented in section 2, electret patterning is primordial to develop efficient and viable eVEH. Various methods from the state of the art to make stable patterned electrets in polymers and $SiO_2$-based layers are presented hereafter.

    a. Polymers

The problem of polymer electrets patterning has been solved for quite a long time [16, 49]. In fact, it has been proven that it is possible to develop stable patterned electrets in CYTOP by etching the electret layer before charging, as presented in Figure 33 [10]. The patterning size is in the order of 100µm.

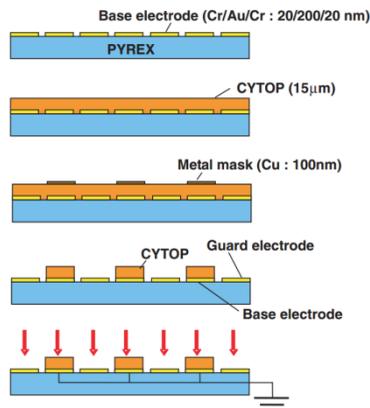

Figure 33. CYTOP electret patterning [10]

Electrostatic Conversion for Vibration Energy Harvesting

Equivalent results have been observed on Teflon AF [16].

However, making patterned $SiO_2$-based electrets is generally more complicated, leading to a strong charge decay and therefore an extremely weak stability.

b. $SiO_2$-based electrets

In fact, an obvious patterning of electret layers would consist in taking full sheet $SiO_2$-based electrets (that have an excellent stability) and by etching them, like it is done on polymer electrets (Figure 34).

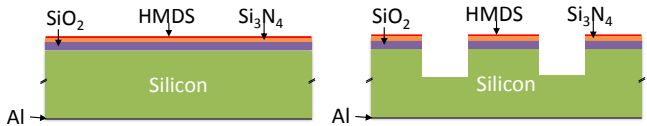

Figure 34. Obvious patterning that does not work

Unfortunately, this obvious patterning does not work because it makes the electret hard to charge (ions or electrons go directly in the silicon wafer) and the stability of these electrets is not good [37]. This is the reason why new and smart solutions have been developed to pattern $SiO_2$-based electrets.

- IMEC

The concept developed by IMEC to make $SiO_2/Si_3N_4$ patterned electrets is based on the observation that a single $SiO_2$ layer is less stable than a superposition of $SiO_2$ and $Si_3N_4$ layers. A drawing of the patterned electrets is provided in Figure 35. This method has been patented by IMEC [50].

These patterned electrets are obtained from a silicon wafer that receives a thermal oxidization to form a $SiO_2$ layer. A $Si_3N_4$ layer is deposited and etched with a patterning. This electret is then charged thanks to a corona discharge. Charges that are not on the $SiO_2/Si_3N_4$ areas are removed thanks to thermal treatments while charges that are on these areas stay trapped inside.

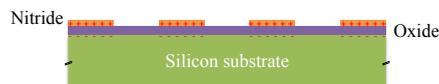

Figure 35. Patterned $SiO_2/Si_3N_4$ electrets from IMEC and used in a structure [45].

The stability of these electrets was proven down to a patterning size of 20μm.

- Sanyo and the University of Tokyo

Naruse et al. [37] developed $SiO_2$ patterned electrets thanks to a different concept. The electret manufacturing process starts with a $SiO_2$ layer on a silicon wafer. The $SiO_2$ layer is metalized with aluminum. The aluminum layer is then patterned and the $SiO_2$ layer is etched as presented in Figure 36. The sample is then charged.

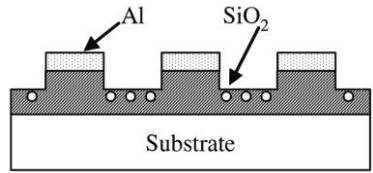

Figure 36. Patterned $SiO_2$ electrets from Sanyo and the University of Tokyo [37]

The guard electrodes of Aluminum form the low surface voltage and the charged areas of $SiO_2$ form the high surface voltage. This potential difference enables to turn mechanical energy into electricity. The hollow structure of $SiO_2$ prevents charge drifting to the guard electrode.

- CEA-LETI [51]

Contrary to the previous methods, the goal of this electret patterning method is to make continuous electret layers. Actually, instead of patterning the electret, it is the substrate of the electret (the silicon wafer) that is patterned thanks to a Deep Reactive Ion Etching (DRIE). The fabrication process of these patterned electrets is similar to the one of full sheet electrets in order to keep equivalent behaviors and above all equivalent stabilities.

The main difference between the two processes (full-sheet and patterned electrets) is the DRIE step that is used to geometrically pattern the electret. The main manufacturing steps are presented in Figure 37. The process starts with a standard p-doped silicon wafer (a). After a lithography step, the silicon wafer is etched by DRIE (b) and cleaned. Wafers are then oxidized to form a 1µm-thick $SiO_2$ layer (c). $SiO_2$ layer on the rear face is then removed by HF while front face is protected by a resin. A 100nm-thick LPCVD $Si_3N_4$ is deposited on the front face (d). Wafers receive a thermal treatment (450°C during 2 hours into $N_2$) and a surface treatment (vapour HMDS) (e). Dielectric layers are then charged by a standard corona discharge to turn them into electrets (f).

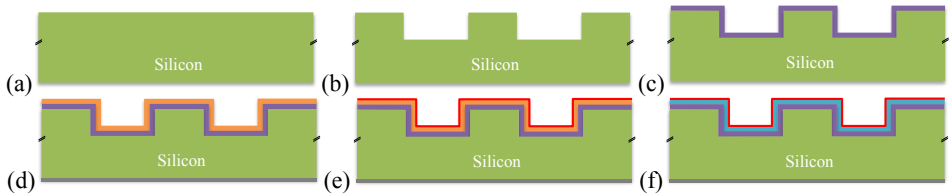

Figure 37. Fabrication process of CEA-LETI's DRIE-patterned electrets

Manufacturing results are presented in Figure 38 for a (e,b,h)=(100µm, 100µm, 100µm) electret (Figure 38(a)). SEM images in Figure 38(b, c, d) show the patterning of the samples and the different constitutive layers. It is interesting to note the continuity of the electret layer even on the right angle in Figure 38(d). The long-term stability of these patterned electrets has been proven thanks to various surface potential decays measurements.

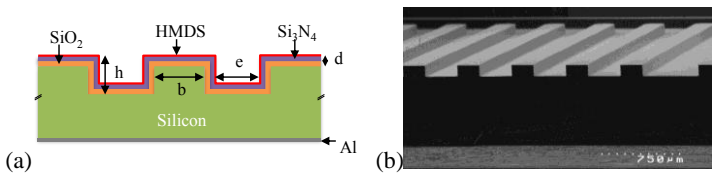

(a)                                          (b)

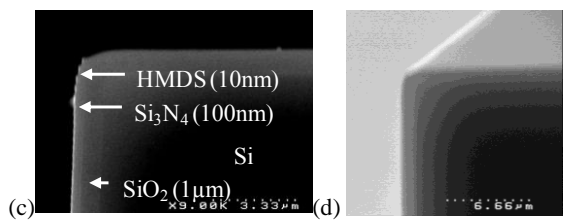

Figure 38. Patterned $SiO_2/Si_3N_4$ electrets from CEA-LETI [51]

We have presented in this section several prototypes of electrostatic VEH and their output powers that may reach some tens or even hundreds of microwatts. This is in agreement with WSN' power needs. Yet, the output voltages are not appropriate for supplying electronic devices as is. This is the reason why a power converter is required.

That power management unit is essential for Wireless Sensor Nodes; this is the topic of the next section.

## 4. Power Management Control Circuits (PMCC) dedicated to electrostatic VEH (eVEH)

The next section is aimed at presenting some examples of PMCC for electrostatic VEH.

### 4.1. Need for Power Management Control Circuit (PMCC)

As presented in section 3, electrostatic vibration energy harvesters are characterized by a high output voltage that may reach some hundreds of volts and a low output current (some 100nA). Obviously, it is impossible to power any application, any electronic device with such a supply source. This is the reason why a power converter and an energetic buffer are needed to develop autonomous sensors. Figure 39 presents the conversion chain.

Power Management Control Circuits (PMCC) can have many functions: changing eVEH resonant frequency, controlling measurement cycles... here, we focus on the power converter and on its control circuit.

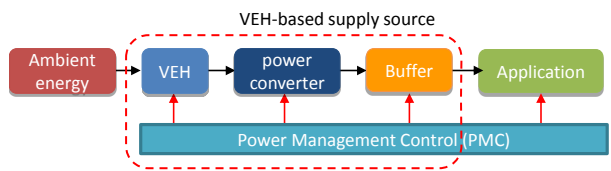

Figure 39. Power Management Control Circuit to develop viable VEH

As eVEH output powers are low (generally <100μW), Power Management Control Circuit must be simple and above all low power. For example, it is difficult to supply a MMPT (Maximum Power Point Tracker) circuits and the number of transistors and operations must be highly limited. We present in the next subsections some examples of Power Management Control Circuit for electret-free and electret-based eVEH.

### 4.2. PMCC for electret-free electrostatic VEH

As the mechanical-to-electrical conversion is not direct, electret-free eVEH need a PMCC able to charge and to discharge the capacitor at the right time. Once more, we will focus on voltage-constrained and charge-constrained cycles.

#### 4.2.1. Voltage-constrained cycles

The voltage-constrained cycle is not often used, and no specific example is available. Yet, Figure 40 presents an example of a PMCC to implement voltage-constrained cycles on electret-free electrostatic converters.

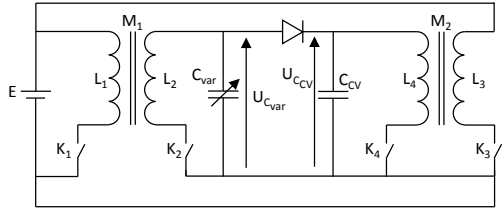

Figure 40. Example of a PMCC to implement voltage-constrained cycles

When the electrostatic converter's capacitance reaches its maximum, a quantity of energy is transferred from the electrical energetic buffer E and stored in the magnetic core $M_1$ by closing $K_1$ during few µs, negligible compared to the mechanical period. This energy is then transferred to the variable capacitor $C_{var}$ by closing $K_2$ during few µs. The voltage $U_{Cvar}$ through $C_{var}$ reaches $U_{CV}$, the constant voltage. The mechanical movement induces a decrease of the electrostatic structure's capacitance $C_{var}$ and a charge $\Delta Q = U_{CV} \Delta C_{var}$ is transferred to the constant voltage storage $C_{CV}$. To maintain $U_{CV}$ approximately constant, a second electrical converter is used. When $U_{CV}$ becomes higher than a threshold voltage, a quantity of energy is transferred from the constant voltage capacitor $C_{CV}$ to the energetic buffer E by closing $K_4$ and then $K_3$. Finally, when the electrostatic structure's capacitance reaches its minimum, the remaining energy stored in the electrostatic structure $C_{var}$ is sent to the energetic buffer by closing $K_2$ and then $K_1$.

Even though this PMCC works, it nevertheless requires two electrical converters that cost in price, space, losses and complexity. In order to have only one electrical converter, the MIT proposed in 2005 the following structure that applies a partial constant voltage cycle [26]:

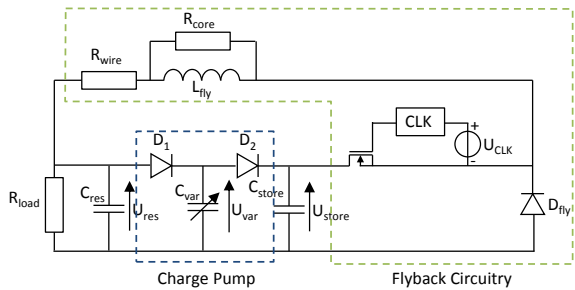

Figure 41. Example of a PMCC from MIT for voltage-constrained cycles [26]

Electrostatic Conversion for Vibration Energy Harvesting

This electronic circuit keeps the electrostatic structure's voltage between two values ($V_{res}$ and $V_{store}$). When the electrostatic structure's capacitance $C_{var}$ increases, its voltage decreases and finishes to reach the low voltage storage $U_{res}$. Then, diode $D_1$ becomes conductive and a current is transferred from $C_{res}$ (storage capacitor) to the electrostatic device. When $C_{var}$'s capacitance decreases, its voltage increases and finally reaches the high voltage storage $U_{store}$. Then diode $D_2$ becomes conductive and a current is transferred from the electrostatic structure to $C_{store}$. This structure works as a charge pump from $C_{res}$ to $C_{store}$. And, in order to close the cycle, one part of the energy transferred to $C_{store}$ is transferred to $C_{res}$ by using an inductive electrical converter. Although this structure uses only one inductive component, it requires a complex electronic circuit to drive the floating transistor connected to the high voltage.

Finally, the constant voltage cycle is not frequently used due to the complex electronic circuits associated.

### 4.2.2. Charge-constrained cycles

Charge-constrained cycles are easier to implement than voltage-constrained cycles as the conversion consists in charging the capacitor when the capacitance is maximal and to let it in open-circuit till it reaches its minimum. On the minimal capacitance, corresponding to the maximal voltage, charges are collected from the converter.

Usually, to reach a high conversion power density, the capacitor must be polarized at a high voltage ($V_1$>100V). Yet, in autonomous devices, only 3V supply sources are available: a first DC-to-DC converter (step-up) is therefore needed to polarize the capacitor at a high voltage (step 1). In the same way, the output voltage on the capacitor after the mechanical-to-electrical conversion (step 2) may reach several hundreds of volts ($V_2$>200-300V) and is therefore not directly usable to power an application: a second converter (step-down) is then necessary (step 3). Obviously, to limit the number of sources, it is interesting to use the same 3V-supply source to charge the electrostatic structure and to collect the charges at the end of the mechanical-to-electrical conversion. Figure 42 sums up the 3-steps conversion process with the two DC-to-DC conversions (DC-to-DC converters) and the mechanical-to-electrical conversion (energy harvester).

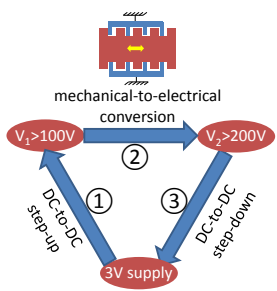

Figure 42. DC-to-DC conversions needed to develop an operational electret-free electrostatic converter and conversion steps

Furthermore, in order to limit the size and the cost of the power converters and the power management control circuit, it is worth combining the step-up and the step-down converters into a single DC-to-DC converter: a bidirectional converter is then used. The two most well-known bidirectional converters are the buck-boost and the flyback converters.

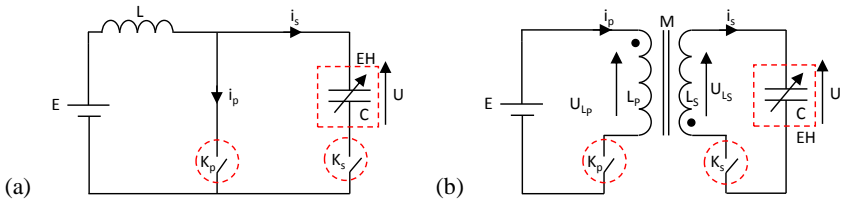

Figure 43. Bidirectional DC-to-DC converters (a) buck-boost and (b) flyback

c.   Bidirectional buck-boost converter

The operating principle of the bidirectional buck-boost converter (Figure 43(a)) is summed up below:

<u>Step 1- Capacitor charging</u>

$K_p$ is closed for a time $t_1$. The energy $E_c$, that has to be sent to the energy harvester to polarize it, is transferred from the supply source E to the inductance L.

$K_p$ is open, and $K_s$ is closed till current $i_s$ becomes equal to 0, corresponding to the time needed to transfer the energy stored in inductance L to the capacitor of the energy harvester C.

<u>Step 2 – Mechanical-to-electrical conversion step</u>

$K_p$ and $K_s$ are open to let the electrostatic converter in open circuit so that the voltage across C may vary freely.

<u>Step 3- Capacitor discharging</u>

$K_s$ is closed for a time $t_2$, to transfer the energy stored in the capacitor C to inductance L and the storage element E.

$K_s$ is open and $K_p$ is closed till $i_p$ becomes equal to 0 corresponding to the time needed to transfer the energy stored in L to the storage element E.

The waveforms of currents in buck–boost converters are presented in Figure 44.

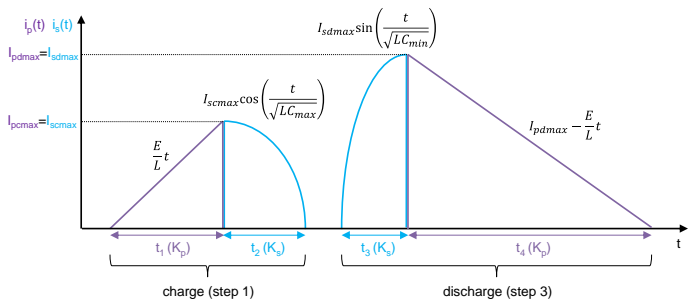

Figure 44. Waveforms of currents in buck–boost converters

This converter has a good conversion efficiency that can reach up to 80-90%. Yet, Flyback converters are generally more suitable for electrostatic energy harvesters where conversion ratios are higher than 30.

d.   Bidirectional flyback converter

Electrostatic Conversion for Vibration Energy Harvesting

The operating principle of the bidirectional flyback converter (Figure 43(b)) is summed up below:

Step 1- Capacitor charging

$K_p$ is closed for a time $t_1$. The energy $E_c$, that has to be sent to the energy harvester to polarize it, is transferred from the supply source E to the inductance $L_p$ that charges the magnetic core M.

$K_p$ is open, and $K_s$ is closed till current $i_s$ becomes equal to 0, corresponding to the time needed to transfer the energy stored in the magnetic core M to the capacitor of the energy harvester C.

Step 2 – Mechanical-to-electrical conversion step

$K_p$ and $K_s$ are open to let the energy harvester in open circuit so that the voltage across C may vary freely.

Step 3- Capacitor discharging

$K_s$ is closed for a time $t_2$, to transfer the energy stored in the capacitor C to the magnetic core M through $L_s$.

$K_s$ is open and $K_p$ is closed till $i_p$ becomes equal to 0 corresponding to the time needed to transfer the energy stored in the magnetic core M to the storage element E.

The waveforms of currents in flyback converters are presented in Figure 45.

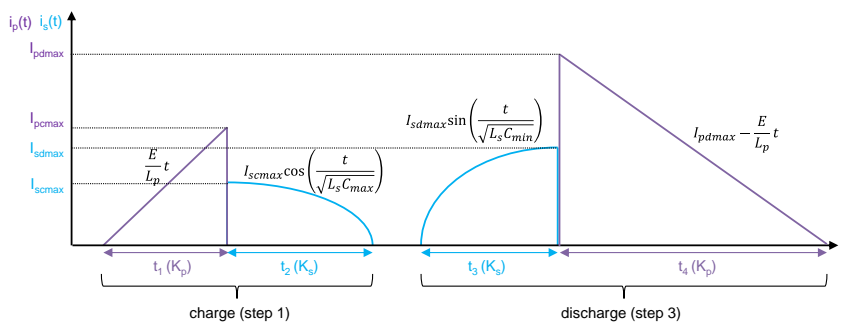

Figure 45. Waveforms of currents in flyback converters

Contrary to buck-boost converters, flyback converters do not need bidirectional transistors ($K_s$ must be bidirectional in buck-boost converters) that complicate the power management circuit and increase losses. Moreover, flyback converters enable to optimize both the windings for the high voltages and the low voltages (while buck-boost converters have only one winding).

These two DC-to-DC conversions (step-up and step-down) can be simplified by using electret-based devices. The next sub-section is focused on the power converters and the power management control circuits for these energy harvesters.

### 4.3. PMCC for Electret-Based Electrostatic VEH

Electret-based eVEH enable to have a direct mechanical-to-electrical conversion without needing any cycles of charges and discharges. As a consequence, it is possible to imagine two kinds of power converters.

### 4.3.1. Passive power converters

Passive power converters are the easiest way to turn the AC high-voltage low-current eVEH output into a 3V DC supply source for WSN. An example of these circuits is presented in Figure 46(a). It consists in a diode bridge and a capacitor that stores the energy from the eVEH.

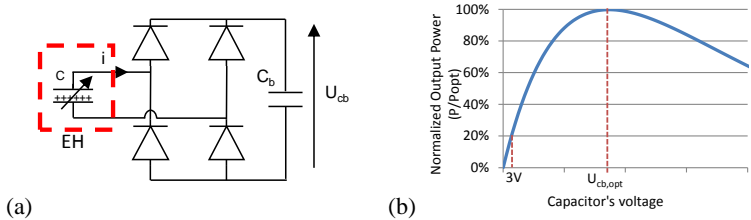

(a)                                                                          (b)

Figure 46. (a) Simple passive power converter – diode bridge-capacitor and (b) optimal output voltage on $U_{cb}$

Such a power converter does not need any PMCC as the energy from the energy harvester is directly transferred to the capacitor. This power conversion is quite simple, but the drawback is the poor efficiency.

Actually, to maximize power extraction from an electret-based electrostatic converter, the voltage across $C_b$ must be close to the half of the eVEH's output voltage in open circuit. This optimal value ($U_{cb,opt}$) is generally equal to some tens or hundreds of volts. To power an electronic device, a 3V source is required: this voltage cannot be maintained directly on the capacitor as it greatly reduces the conversion efficiency of the energy harvester (Figure 46(b)).

The solution to increase the efficiency of the energy harvester consists in using active power converters.

### 4.3.2. Active power converters

As eVEH' optimal output voltages are 10 to 100 times higher than 3V, a step-down converter is needed to fill the buffer. The most common step-down converters are the buck, the buck-boost and the flyback converters. We focus here on the flyback converter that gives more design flexibilities (Figure 47).

Many operation modes can be developed to turn the eVEH high output voltages into a 3V supply source. Here, we focus on two examples: (i) energy transfer on maximum voltage detection and (ii) energy transfer with a pre-storage to keep an optimal voltage across the electrostatic converter.

    a.    Energy transfer on a maximum voltage detection

The concept of this power conversion is to send the energy from the energy harvester to the 3V energy buffer when the eVEH output voltage reaches its maximum.

The power management control circuit is aimed at finding the maximum voltage across the energy harvester and to close $K_p$ (Figure 48) to send the energy from the eVEH to the magnetic circuit. Then $K_s$ is closed to send the energy from the magnetic circuit to the buffer $C_b$. The winding ratio m is determined from the voltage ratio between the primary and the secondary.

Electrostatic Conversion for Vibration Energy Harvesting

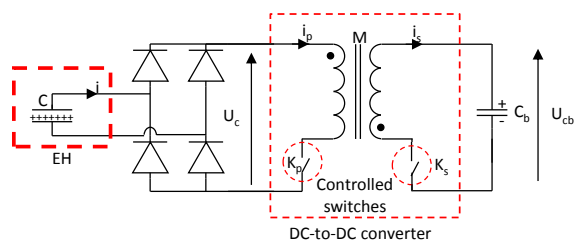

Figure 47. Energy transfer on maximum voltage detection

Figure 48 presents the voltages and the currents on the primary and on the secondary during the power transfer.

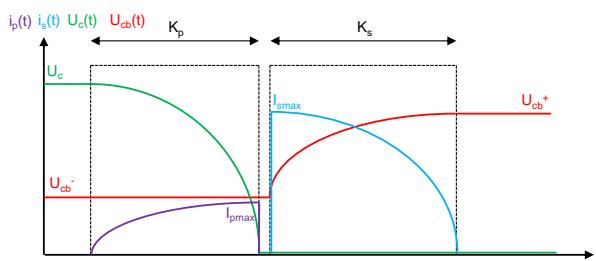

Figure 48. Voltages and currents during power conversion

As eVEH capacitances are quite small, parasitic capacitances of the primary winding may have a strong negative impact on the output powers, increasing conversion losses. An alternative consists in using a pre-storage capacitor.

b. Energy transfer with a pre-storage capacitor

In this operation mode, a pre-storage capacitor $C_p$ is used to store the energy from the eVEH and to maintain an optimal voltage across the diode bridge in order to optimize the energy extraction from the eVEH.

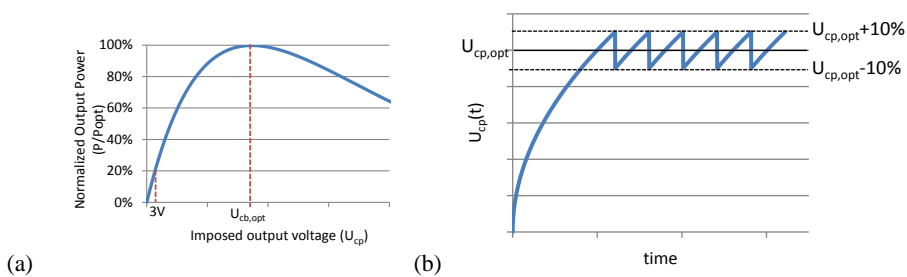

(a)                                   (b)

Figure 49. (a) eVEH output power vs imposed output voltage and (b) $U_{cp}(t)$

The goal of the PMCC is to maintain the voltage quite constant across the diode bridge (+/-10% $U_{cp,opt}$). Then, when $U_{cp}$ reaches $U_{cp,opt}$+10%, one part of the energy stored in $C_p$ is sent to $C_b$ through the flyback converter.

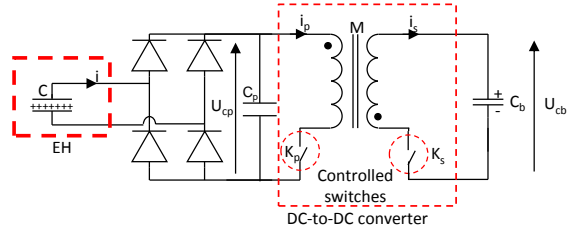

Figure 50. Energy transfer with pre-storage

Voltages and currents during the electrical power transfer are presented in Figure 51.

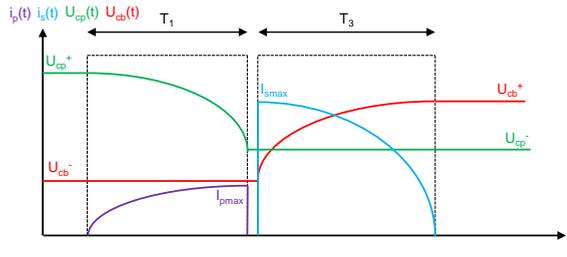

Figure 51. Voltages and currents during power conversion

As $C_p$ can be in the order of some tens to hundreds of nanofarads, transformer's parasitic capacitances have smaller impacts on eVEH's output power.

This power conversion principle also enables to use multiple energy harvesters in parallel with only one transformer and above all only one PMCC (which is not the case with the maximum voltage detection).

We have presented some examples of power converters able to turn the raw output powers of the energy harvesters into supply sources able to power electronic devices. Thanks to this, and low power consumptions of WSN' nodes, it is possible to develop autonomous wireless sensors using the energy from vibrations from now on. The last section gives an assessment of this study.

## 5.  Assessments & Perspectives

In this last section, we present our vision of eVEH and their perspectives for the future.

### 5.1. Assessments

Electrostatic VEH are doubtless the less known vibration energy harvesters, and especially compared to piezoelectric devices. Yet, these devices have undeniable advantages: the possibility to develop structures with high mechanical-to-electrical couplings, to decouple the mechanical-to-mechanical converter and the mechanical-to-electrical converter, to develop low-cost devices able to withstand high temperatures…

Moreover, even if these devices have incontestable drawbacks as well, such as low capacitances, high output voltages and low output currents, it has been proven that they can be compatible with WSN needs as soon as a power converter is inserted between the VEH and the device to supply.

## 5.2. Limits

Obviously, eVEH have drawbacks and limitations. We present in this subsection the four most important limits of these devices.

i.    Integration of devices. The question of size reduction is common to all VEH. Actually, as the output power is proportional to the mobile mass, it is not necessarily useful to reduce VEH' dimensions at any cost. Moreover, it becomes particularly difficult to design devices with a resonant frequency lower than 50Hz when working with small-scale devices. As a consequence, to have a decent output power (>10µW) and a robust device, it is hard to imagine devices smaller than 1cm².

ii.   Working frequency and frequency bandwidth. Ambient vibrations are characterized by a low frequency, generally lower than 100Hz. Moreover, when looking at the vibrations spectra, it appears that they are spread over a wide frequency range. This implies that we need to develop low-frequency broadband devices; this may rise to many problems in the design and the manufacturing of the springs. Indeed, to build low-frequency devices, especially with small-scale devices, thin and long guide beams are needed. They are particularly fragile and are moreover submitted to high strains and stresses.

iii.  Gap control. eVEH output powers are greatly linked to the capacitance variations, that must be maximized. Therefore, the air gap must be controlled precisely and minimized to reach high capacitances. Yet, it is also important to take care of pull-in and electrical breakdown problems.

iv.   Electret stability. Electret stability may also be critical. Actually, electret stability is strongly linked to environmental conditions, for example to humidity and temperature. Moreover, contacts between electrets and electrodes must be avoided as they generally lead to breakdown and discharge of electrets.

## 5.3. Perspectives

Like all VEH (piezoelectric, electromagnetic or electrostatic), the most critical point to improve is the frequency bandwidth that must be largely increased to develop viable and adaptable devices.

Indeed, a wide frequency bandwidth is firstly necessary to develop robust devices. VEH are submitted to a large amount of cycles (16 billion cycles for a device that works at 50Hz during 10 years), that may change the resonant frequency of the energy harvester due to fatigue. Then, the energy harvester's resonant frequency is not tuned to the ambient vibrations' frequency anymore. Therefore, it is absolutely primordial to develop devices able to maintain their resonant frequency equal to the vibration frequency.

Wideband energy harvesters are also interesting to develop adaptable devices, able to work in many environments and simple to set up and to use. There is a real need for Plug and Play devices.

Figure 52 presents our vision of VEH today: VEH market as a function of the time and the two technological bottlenecks linked to working frequency bandwidths. In our opinion, today's VEH are yet suitable for industry; increasing working frequency bandwidths and developing plug and play devices are the only way to conquer new markets.

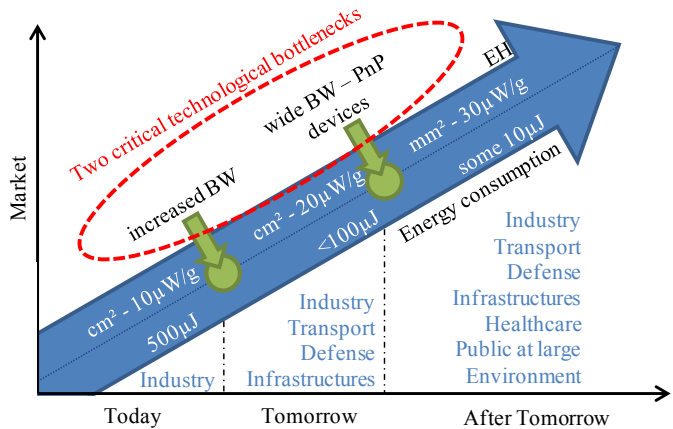

Figure 52. Vibration Energy Harvesters – Perspectives [52]

## 6. Conclusions

We have presented in this chapter the basic concepts and theories of electrostatic converters and electrostatic vibration energy harvesters together with some prototypes from the state of the art, adopting a "global system" vision.

Electrostatic VEH are increasingly studied from the early 2000s. Unfortunately, no commercial solution is on the market today, dedicating these devices to research.

We believe that this is a pity because they have undeniable advantages compared to piezoelectric or electromagnetic devices. The first in importance is probably the possibility to manufacture low cost devices (low cost and standard materials). Obviously, the limited frequency bandwidth of vibration energy harvesters does not help the deployment of these devices, even if some solutions are currently under investigation. Yet, with this increasing need to get more information from our surroundings, we can expect that these systems will match industrial needs and find industrial applications.

Anyway, electrostatic converters and electrostatic vibration energy harvesters remain an interesting research topic that gathers material research (electrets), power conversion, low consumption electronics, mechanics and so on…

### References
[1] Boisseau S, Despesse G. Energy harvesting, wireless sensor networks & opportunities for industrial applications. EE Times 2012.
http://www.eetimes.com/design/smart-energy-design/4237022/Energy-harvesting--wireless-sensor-networks---opportunities-for-industrial-applications
[2] Williams C B, Yates R B. Analysis Of A Micro-electric Generator For Microsystems. Proc. Eurosensors 1995;1: 369-72.
[3] Eguchi M. On the Permanent Electret. Philosophical Magazine 1925; 49:178.
[4] Kotrappa P. Long term stability of electrets used in electret ion chambers. Journal of Electrostatics 2008;66: 407-9.
[5] Kressmann R, Sessler G, Gunther P. Space-charge electrets. Transactions on Dielectrics and Electrical Insulation 1996;3: 607-23.

[6] Gunther P, Ding H, Gerhard-Multhaupt R. Electret properties of spin-coated Teflon-AF films. Proc. Electrical Insulation and Dielectric Phenomena 1993: 197-202.

[7] Sessler G. Electrets [3rd Edition] [in Two Volumes]. Laplacian Press, Morgan Hill, 1999.

[8] Gunther P. Charging, long-term stability and TSD measurements of SiO2 electrets. Transactions on Electrical Insulation 1989;24(3): 439-42.

[9] Leonov V, Fiorini P, Van Hoof C. Stabilization of positive charge in SiO2/Si3N4 electrets. IEEE transactions on dielectrics and electrical insulation. 2006;13(5): 1049-56.

[10] Sakane Y, Suzuki Y, Kasagi N. The development of a high-performance perfluorinated polymer electret and its application to micro power generation. IOP Journal of Micromechanics and Microengineering 2008;18(104011). http://dx.doi.org/10.1088/0960-1317/18/10/104011

[11] Rychkov D, Gerhard R. Stabilization of positive charge on polytetrafluoroethylene electret films treated with titanium-tetrachloride vapor. Appl. Phys. Lett. 2011;98(122901).

[12] Schwödiauer R, Neugschwandtner G, Bauer-Gogonea S, Bauer S, Rosenmayer T. Dielectric and electret properties of nanoemulsion spin-on polytetrafluoroethylene films. Appl. Phys. Lett. 2000;76(2612).

[13] Kashiwagi K, Okano K, Morizawa Y, Suzuki Y. Nano-cluster-enhanced High-performance Perfluoro-polymer Electrets for Micro Power Generation. Proc. PowerMEMS 2010:169-72.

[14] Kashiwagi K, Okano K, Miyajima T, Sera Y, Tanabe N, Morizawa Y, Suzuki Y. Nano-cluster-enhanced High-performance Perfluoro-polymer Electrets for Micro Power Generation. IOP J. Micromech. Microeng.2011;21(125016). http://dx.doi.org/10.1088/0960-1317/21/12/125016

[15] Boisseau S, Despesse G, Ricart T, Defay E, Sylvestre A. Cantilever-based electret energy harvesters. IOP Smart Materials and Structures 2011; 20(105013). http://dx.doi.org/10.1088/0964-1726/20/10/105013

[16] Boland J, Chao Y, Suzuki Y, Tai Y. Micro electret power generator. Proc. MEMS 2003:538-41.

[17] Boisseau S, Despesse G, Sylvestre A. Optimization of an electret-based energy harvester. Smart Materials and Structures 2010;19(075015). http://dx.doi.org/10.1088/0964-1726/19/7/075015

[18] Meninger S, Mur-Miranda J O, Amirtharajah R, Chandrakasan A, Lang J. Vibration-to-electric energy conversion. IEEE transactions on very large scale integration (VLSI) 2011;9(1): 64-75.

[19] Tashiro R, Kabei N, Katayama K, Tsuboi E, Tsuchiya K. Development of an electrostatic generator for a cardiac pacemaker that harnesses the ventricular wall motion. Journal of Artificial Organs 2002;5:239-45.

[20] Roundy S. Energy Scavenging for Wireless Sensor Nodes with a Focus on Vibration to Electricity Conversion. PhD Thesis. University of California, Berkeley, 2003.

[21] Despesse G, Chaillout J J, Jager T, Léger J M, Vassilev A, Basrour S, Charlot B. High damping electrostatic system for vibration energy scavenging. Proc. sOc-EUSAI 2005:283-6.

[22] Basset P, Galayko D, Paracha A, Marty F, Dudka A, Bourouina T. A batch-fabricated and electret-free silicon electrostatic vibration energy harvester. IOP Journal of Micromechanics and Microengineering 2009;19(115025). http://dx.doi.org/10.1088/0960-1317/19/11/115025

[23] Hoffmann D, Folkmer B, Manoli Y. Fabrication and characterization of electrostatic micro-generators. Proc. PowerMEMS 2008: 15.

[24] Roundy S. Energy Scavenging for Wireless Sensor Networks with Special Focus on Vibrations. Hardcover, Springer, 2003.

[25] Mitcheson P, Green T C, Yeatmann E M, Holmes A S. Architectures for vibration-driven micropower generators. J. of Microelect. Systems 2004;13: 429-40.

[26] Chih-Hsun Yen B, Lang J. A variable capacitance vibration-to-electric energy harvester. IEEE Trans. Circuits Syst. 2006;53: 288-95.

[27] Jefimenko O, Walker D K. Electrostatic Current Generator Having a Disk Electret as an Active Element. Transactions on Industry Applications 1978;IA-14: 537-40.

[28] Tada Y. Experimental Characteristics of Electret Generator, Using Polymer Film Electrets. Japanese Journal of Applied Physics 1992;31: 846-51.

[29] Genda T, Tanaka S, Esashi M. High power electrostatic motor and generator using electrets. Proc. Transducers 2003;1: 492-5.

[30] Boland J. Micro electret power generators. PhD thesis. California Institute of Technology. 2005.

[31] Mizuno M, Chetwynd D. Investigation of a resonance microgenerator. IOP Journal of micromechanics and Microengineering 2003;13: 209-16. http://dx.doi.org/10.1088/0960-1317/13/2/307

[32] Sterken T, Fiorini P, Altena G, Van Hoof C, Puers R. Harvesting Energy from Vibrations by a Micromachined Electret Generator. Proc. Transducers 2007: 129-32.

[33] Tsutsumino T, Suzuki Y, Kasagi N, Sakane Y. Seismic Power Generator Using High-Performance Polymer Electret. Proc. MEMS 2006: 98-101.

[34] Lo H W, Tai Y C, Parylene-HT-based electret rotor generator. Proc. MEMS 2008:984-7.

[35] Zhang X, Sessler G M. Charge dynamics in silicon nitride/silicon oxide double layers. Applied Physics Letters 2001;78: 2757-9.

[36] Edamoto M, Suzuki Y, Kasagi N, Kashiwagi K, Morizawa Y, YokoyamaT, Seki T, Oba M. Low-resonant-frequency micro electret generator for energy harvesting application. Proc. MEMS 2009: 1059–62.

[37] Naruse Y, Matsubara N, Mabuchi K, Izumi M, Suzuki S. Electrostatic micro power generation from low-frequency vibration such as human motion. IOP Journal of Micromechanics and Microengineering 2009;19(094002). http://dx.doi.org/10.1088/0960-1317/19/9/094002

[38] Suzuki Y, Miki D, Edamoto M, Honzumi M. A MEMS Electret Generator with Electrostatic Levitation for Vibration-Driven Energy Harvesting Applications. IOP J. Micromech. Microeng. 2010;20(104002). http://dx.doi.org/10.1088/0960-1317/20/10/104002

[39] Miki D, Honzumi M, Suzuki S, Kasagi N. Large-amplitude MEMS electret generator with nonlinear spring. Proc. MEMS 2010: 176-9.

[40] Suzuki Y, Edamoto M, Kasagi N, Kashwagi K, Morizawa Y. Micro electret energy harvesting device with analogue impedance conversion circuit. Proc. PowerMEMS 2008: 7-10.

[41] Suzuki Y. Recent Progress in MEMS Electret Generator for Energy Harvesting. IEEJ Trans. Electr. Electr. Eng. 2011;6(2): 101-11.

[42] Boland J, Messenger J, Lo K, Tai Y. Arrayed liquid rotor electret power generator systems. Proc. MEMS 2005: 618-21.

[43] Lo H W, Whang R, Tai Y C. A simple micro electret power generator. Proc. MEMS 2007: 859-62.

[44] Yang Z, Wang J, Zhang J. A micro power generator using PECVD SiO2/Si3N4 double layer as electret. Proc. PowerMEMS 2008:317-20.

[45] Halvorsen E, Westby E R, Husa S, Vogl A, Østbø N P, Leonov V, Sterken T, Kvisterøy T. An electrostatic Energy harvester with electret bias. Proc. Transducers 2009:1381–4.

[46] Kloub H, Hoffmann D, Folkmer B and Manoli Y. A micro capacitive vibration Energy harvester for low power electronics. Proc. PowerMEMS 2009:165–8.

[47] Honzumi M, Ueno A, Hagiwara K, Suzuki Y, Tajima T, Kasagi N. Soft-X-Ray-charged vertical electrets and its application to electrostatic transducers. Proc. MEMS 2010:635–8.

[48] Shimizu N. Omron, Sanyo Prototype Mini Vibration-Powered Generators, Nikkei Electronics Asia, Feb 16, 2009. [Online] 2009.
http://techon.nikkeibp.co.jp/article/HONSHI/20090119/164257/.

[49] Suzuki Y, Miki D, Edamoto M, Honzumi M. A MEMS Electret Generator With Electrostatic Levitation For Vibration-Driven Energy Harvesting Applications. IOP J. Micromech. Microeng. 2010;20(104002). http://dx.doi.org/10.1088/0960-1317/20/10/104002

[50] Leonov V. Patterned Electret Structures and Methods for Manufacturing Patterned Electret Structures. Patent 2011/0163615. 2011.

[51] Boisseau S, Duret A B, Chaillout J J, Despesse G. New DRIE-Patterned Electrets for Vibration Energy Harvesting. Proc. European Energy Conference 2012.

[52] Boisseau S, Despesse G. Vibration energy harvesting for wireless sensor networks: Assessments and perspectives. EE Times 2012.
http://www.eetimes.com/design/smart-energy-design/4370888/Vibration-energy-harvesting-for-wireless-sensor-networks--Assessments-and-perspectives

## Aknowledgements

The authors would like to thank their VEH coworkers: J.J. Chaillout, C. Condemine, A.B. Duret, P. Gasnier, J.M. Léger, S. Soubeyrat, S. Riché, S. Dauvé and J. Willemin for their contributions to this chapter.